\documentclass[aps,prb,reprint,superscriptaddress]{revtex4-2}
\usepackage{graphicx}
\usepackage{amsmath}
\usepackage{amssymb}
\usepackage{bm}
\usepackage{physics}
\usepackage{subfigure}
\usepackage[colorlinks=true,linkcolor=blue,urlcolor=blue,citecolor=blue]{hyperref}
\usepackage{times}
\usepackage{changes}

\usepackage{tikz}
\usetikzlibrary{shapes,arrows,fit,positioning}

\begin{document}

\newcommand{\Hop}{\hat{H}}
\newcommand{\Himp}{\hat{H}_{\rm imp}}
\newcommand{\Hint}{\hat{H}_{\rm int}}
\newcommand{\Hs}{\hat{H_{\rm S}}}
\newcommand{\Uimp}{\hat{U}_{\rm imp}}
\newcommand{\gHimp}{\mathcal{H}_{\rm imp}}
\newcommand{\gUimp}{\mathcal{U}_{\rm imp}2}
\newcommand{\Hbath}{\hat{H}_{\rm bath}}
\newcommand{\Hhyb}{\hat{H}_{\rm hyb}}

\newcommand{\sop}{\hat{s}}
\newcommand{\bop}{\hat{b}}
\newcommand{\bdop}{\hat{b}^{\dagger}}
\newcommand{\aop}{\hat{a}}
\newcommand{\adop}{\hat{a}^{\dagger}}
\newcommand{\sgp}{\hat{\sigma}^+}
\newcommand{\sgx}{\hat{\sigma}^x}
\newcommand{\sgy}{\hat{\sigma}^y}
\newcommand{\sgz}{\hat{\sigma}_z}
\newcommand{\nop}{\hat{n}}

\newcommand{\cop}{\hat{c}}
\newcommand{\cdop}{\hat{c}^{\dagger}}
\newcommand{\hc}{{\rm H.c.}}
\newcommand{\rhotot}{\hat{\rho}_{\mathrm{tot}}}
\newcommand{\rhoop}{\hat{\rho}}
\newcommand{\rhoimp}{\hat{\rho}_{\mathrm{imp}}}
\newcommand{\rhobath}{\hat{\rho}_{\mathrm{bath}}}
\newcommand{\Zimp}{Z_{{\rm imp}}}
\newcommand{\Zbath}{Z_{\mathrm{bath}}}
\newcommand{\mea}{\mathcal{D}}
\newcommand{\gK}{\mathcal{K}}
\newcommand{\gI}{\mathcal{I}}
\newcommand{\gF}{\mathcal{F}}
\newcommand{\bolda}{\bm{a}}
\newcommand{\boldabar}{\bar{\bm{a}}}
\newcommand{\abar}{\bar{a}}
\newcommand{\im}{{\rm i}}
\newcommand{\contour}{\mathcal{C}}
\newcommand{\gA}{\mathcal{A}}
\newcommand{\gB}{\mathcal{B}}
\newcommand{\gM}{\mathcal{M}}
\newcommand{\boldeta}{\bm{\eta}}
\newcommand{\boldetabar}{\bar{\bm{\eta}}}
\newcommand{\etabar}{\bar{\eta}}
\newcommand{\parity}{\mathcal{P}}
\newcommand{\current}{\mathcal{J}}
\newcommand{\JW}{{\rm JW}}
\newcommand{\pronyerror}{\varsigma_p}
\newcommand{\WI}{{\rm W}^I}
\newcommand{\WII}{{\rm W}^{II}}
\newcommand{\branch}{\mathcal{C}}

\newcommand{\EqDef}{\stackrel{\mathrm{def}}{=}}

\newcommand{\rem}[1]{{\color{red}{\sout{#1}}}}
\newcommand{\gcc}[1]{{\color{red}#1}}
\definechangesauthor[name=RF,color=orange]{RF}

\title{Solving equilibrium quantum impurity problems on the L-shaped Kadanoff-Baym contour}

\author{Ruofan Chen}
\affiliation{College of Physics and Electronic Engineering, and Center for Computational Sciences, Sichuan Normal University, Chengdu 610068, China}


\author{Chu Guo}
\email{guochu604b@gmail.com}
\affiliation{Key Laboratory of Low-Dimensional Quantum Structures and Quantum Control of Ministry of Education, Department of Physics and Synergetic Innovation Center for Quantum Effects and Applications, Hunan Normal University, Changsha 410081, China}
\date{\today}

\begin{abstract}
The path integral formalism is the building block of many powerful numerical methods for quantum impurity problems. However, existing fermionic path integral based numerical calculations have only been performed in either the imaginary-time or the real-time axis, while the most generic scenario formulated on the L-shaped Kadanoff-Baym contour is left unexplored. In this work, we extended the recently developed Grassmann time-evolving matrix product operator (GTEMPO) method to solve quantum impurity problems directly on the Kadanoff-Baym contour. The resulting method is numerically exact, with only two sources of numerical errors, e.g., the time discretization error and the matrix product state bond truncation error.
The accuracy of this method is numerically demonstrated against exact solutions in the noninteracting case, and against existing calculations on the real- and imaginary-time axes for the single-orbital Anderson impurity model. We also show that the numerical errors of the method can be well suppressed as we refine the hyperparameters. Our method is a perfect benchmarking baseline for its alternatives which often employ less-controlled approximations, and can also be used as a real-time impurity solver in dynamical mean field theory.
\end{abstract}
\maketitle

\section{Introduction}
The quantum impurity problem (QIP), which considers an impurity of a few energy levels that is coupled to a continuous noninteracting bath, represents a cornerstone in condensed matter physics for studying strongly correlated effects~\cite{mahan2000-many} and open quantum effects~\cite{weiss1993-quantum}. It is also the building block in quantum embedding methods such as the dynamical mean field theory (DMFT)~\cite{georges1996-dynamical,GullWerner2011}.

The observables of interest in solving QIPs are the multi-time correlations of the impurity.
Existing numerical approaches for QIPs can be primarily divided into two categories: (1) the wave-function based approach and (2) the path integral (PI) based approach. In the first category, one discretizes the bath into a finite number of modes and parameterizes the impurity-bath wave function with some ansatz, then one can compute the multi-time impurity correlations by performing real-time evolution of the wave function ansatz. The representative methods in the first category include the exact diagonalization~\cite{CaffarelKrauth1994,KochGunnarsson2008,GranathStrand2012,LuHaverkort2014,ZaeraLin2020,HeLu2014,HeLu2015}, numerical renormalization group~\cite{Wilson1975,Bulla1999,BullaPruschke2008,Frithjof2008,ZitkoPruschke2009,DengGeorges2013,StadlerWeichselbaum2015,LeeWeichselbaum2016,LeeWeichselbaum2017} and matrix product state (MPS) based methods~\cite{WolfSchollwock2014b,GanahlEvertz2014,GanahlVerstraete2015,WolfSchollwock2015,GarciaRozenberg2004,NishimotoJeckelmann2006,WeichselbaumDelft2009,BauernfeindEvertz2017,LuHaverkort2019,WernerArrigoni2023,KohnSantoro2021,KohnSantoro2022}. The scalability or accuracy of these methods are basically limited by the discretization of the continuous bath.

In the second category, the starting point is the PI of the impurity, where the bath has already been integrated out analytically via the Feynman-Vernon influence functional (IF)~\cite{FeynmanVernon1963} and one is only left with the impurity degrees of freedom in the temporal domain. Most outstandingly, the class of continuous-time Quantum Monte Carlo (CTQMC) methods directly draw samples from the perturbative expansion of the PI, which could efficiently yield exact results in the imaginary-time axis~\cite{GullWerner2011,RubtsovLichtenstein2005,GullTroyer2008,WernerMillis2006b,WernerMillis2006,ShinaokaWerner2017,EidelsteinCohen2020}. QMC calculations on the real-time axis are not as successful, but have been improved significantly in recent years with advanced time-domain extrapolation techniques to tame the dynamical sign problem, such as the inchworm QMC~\cite{CohenGull2014,CohenMillis2014,CohenMillis2015,ChenReichman2017a,ChenReichman2017b,ErpenbeckCohen2023}. The hierarchical equation of motion (HEOM) method which truncates the coupled operator equations to a certain order~\cite{YoshitakaKubo1989,jin2007-dynamics,jin2008-exact}, is mostly applied on the real-time axis up to date~\cite{yan2016-dissipation,cao2023-recent}. There are also direct perturbative calculations on the L-shaped contour~\cite{strand2015-beyond,dong2022-excitations}.
Another recently emerging class of methods in the second category explores the MPS representation of the multi-time impurity dynamics, which is completely different from conventional wave-function based MPS methods. The idea is to either represent the Feynman-Vernon IF in the Fock space as a fermionic MPS (which will be referred to as the tensor network IF method afterwards)~\cite{ThoennissAbanin2023a,ThoennissAbanin2023b,NgReichman2023}, or directly represent the integrand of the PI in the coherent state basis as a Grassmann MPS (GMPS)~\cite{ChenGuo2024a} which is referred to as the Grassmann time-evolving matrix product operator (GTEMPO) method (We note that GTEMPO is an extension of the time-evolving matrix product operator method~\cite{StrathearnLovett2018} for bosonic impurity problems to the fermionic case). 
Since GTEMPO directly works in the coherent state basis (which is the defining basis of the fermionic PI), the existing analytical expressions of the Feynman-Vernon IF can be readily used (this is in contrast with the tensor network IF method where one needs to translate the these expressions into the Fock space which could be nontrivial, especially on the L-shaped contour).
Up to date these two strategies have been applied for both the real-time~\cite{ThoennissAbanin2023a,ThoennissAbanin2023b,NgReichman2023,ChenGuo2024a,ChenGuo2024c,GuoChen2024d} and imaginary-time calculations~\cite{KlossAbanin2023,ChenGuo2024b}.

In this work we extend the GTEMPO method to directly solve QIPs on the L-shaped Kadanoff-Baym contour, which is referred to as the mixed-time GTEMPO method afterwards to distinguish it with previous GTEMPO methods on the real- and imaginary-time axes. 
We note that in existing PI-based approaches, the equilibrium retarded Green's function is either calculated by analytic continuation from the Matsubara Green's function on the imaginary-time axis~\cite{jarrell1989-dynamical,FeiGull2021}, or by performing real-time evolution from a separable impurity-bath initial state till (approximately) infinite time and calculating Green's functions afterwards~\cite{ChenGuo2024c,GuoChen2024e}. The first approach is known to be numerically ill-posed~\cite{WolfSchollwock2015,FeiGull2021}, while the second approach relies on the quality of the equilibrium state which is heuristic in general. We note that Ref.~\cite{GuoChen2024e} directly aims at the infinite-time limit by using infinite MPS techniques, which is in principle free of the error from the equilibration dynamics, however, it uses more hyperparameters than the approach in this work and only contains information on the real-time axis. By performing the mixed-time GTEMPO calculation directly on the Kadanoff-Baym contour, we can avoid all these uncontrolled approximations. The only two sources of errors in this approach, as similar to the real-time and imaginary-time GTEMPO methods, are the time discretization error and the MPS bond truncation error. 
We demonstrate the accuracy of this method against exact solutions in the noninteracting case, and against the real- and imaginary-time GTEMPO calculations in the single-orbital Anderson impurity model (AIM). 
We also show that the numerical errors of this method can be well suppressed by refining the hyperparameters.
Importantly, we observe that we can obtain accurate results with a bond dimension which is not larger than that required in the imaginary-time calculations. The mixed-time GTEMPO method can be a perfect benchmarking baseline for accessing the accuracy of alternative numerical approaches with less-controlled approximations, and can also be used as a real-time impurity solver in DMFT.

\section{Method description}

\subsection{The model}
\label{subsec:model}
For notational briefness, we will present our method based on the single-orbital Anderson impurity model which describes the interaction between a single localized electron and a noninteracting bath of itinerant electrons. The total Hamiltonian can be written as
\begin{align}
\Hop = \Himp + \Hint,
\end{align}
where $\Himp$ is the impurity Hamiltonian:
\begin{align}
\Himp = (\varepsilon_d-\frac{1}{2}U)\sum_{\sigma}\adop_{\sigma}\aop_{\sigma} + U \adop_{\uparrow}\aop_{\uparrow}\adop_{\downarrow}\aop_{\downarrow},
\end{align}
with $\varepsilon_d$ the on-site energy of the impurity and $U$ the Coulomb interaction. $\Hint$ contains the free bath Hamiltonian and the coupling between the impurity and the bath:
\begin{align}
\Hint = \sum_{k, \sigma}\varepsilon_k \cdop_{k, \sigma}\cop_{k, \sigma} + \sum_{k, \sigma}\left(V_k \adop_{\sigma}\cop_{k, \sigma} + \hc \right),
\end{align}
where $\varepsilon_k$ is the band energy and $V_k$ is the coupling strength. The effects of $\Hint$ on the impurity dynamics is completely characterized by the bath spectrum density $J(\varepsilon) = \sum_k V_k^2 \delta(\varepsilon - \varepsilon_k)$.


For quantum impurity problems, the multi-time correlation functions of the impurity, evaluated with respect to the impurity-bath equilibrium state, are often the central quantities of interest. Generally, these correlations describe the response of the impurity plus bath in equilibrium as a whole to an external perturbation performed on the impurity. The single-particle retarded Green's function, in particular, which is a two-time impurity correlation function, is also the central quantity to calculate in DMFT.

\subsection{The path integral formalism}
\label{subsec:path-integral}

\begin{figure}[htbp]
  \centerline{\includegraphics[]{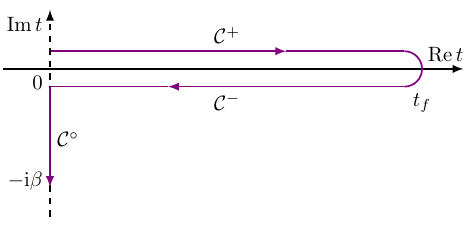}}
  \caption{The L-shaped Kadanoff-Baym contour $\branch=\branch^+\cup \branch^-\cup \branch^{\circ}$, where the arrows indicate the contour ordering.}
  \label{fig:kadanoff}
\end{figure}

In the PI formalism, the multi-time impurity correlations can be naturally evaluated on a L-shaped contour, which can be seen as follows. 
Starting from the impurity-bath equilibrium state $e^{-\beta\Hop}$ with inverse temperature $\beta$, the impurity-bath state at time $t_f$ is 
\begin{align}
\rhoop(t_f)=e^{-\im\Hop t_f}e^{-\beta\Hop}e^{\im\Hop t_f}.
\end{align}
The impurity partition function at time $t_f$ is defined as 
\begin{align}
  \label{eq:Zimp}
  \Zimp(t_f) &\EqDef \Zbath^{-1}\Tr \rhoop(t_f) \nonumber \\ 
  &=\Zbath^{-1}\Tr[e^{-\beta\Hop}e^{\im\Hop t_f}e^{-\im\Hop t_f}],
\end{align}
where $\Zbath$ is the free bath partition function, and we have used the cyclic property of the trace in the second line.
If we read the term inside the square bracket in the second line of Eq.(\ref{eq:Zimp}) from right to left, we can think of the evolution as along such a contour: it starts from time $0$ to $t_f$ by a forward evolution $e^{-\im\Hop t_f}$,  then it returns back  to time 0 by a backward evolution $e^{\im\Hop t_f}$, and finally it goes to $-\im\beta$ by an imaginary time evolution $e^{-\beta \Hop}$. The whole contour, denoted as $\branch$ in this work, is usually referred to as the Kadanoff-Baym contour~\cite{kadanoff1962-quantum,AokiWerner2014}. $\branch$ consists of three branches: the \textit{forward branch} $\branch^+:0\to t_f$, the \textit{backward branch} $\branch^-:t_f\to0$, and the \textit{imaginary branch} $\branch^{\circ}:0\to-\im\beta$, as shown in Fig.~\ref{fig:kadanoff}.

The PI for Eq.(\ref{eq:Zimp}) can be formally written as
\begin{align}\label{eq:PI}
\Zimp(t_f) = \int \mathcal{D}[\boldabar,\bolda] \gK\left[\boldabar, \bolda \right]\prod_{\sigma}\gI_{\sigma}\left[\boldabar_{\sigma}, \bolda_{\sigma}\right],
\end{align}
where  $\boldabar_{\sigma} = \{\abar_{\sigma}(t)\}$, $\bolda_{\sigma} = \{a_{\sigma}(t)\}$ are Grassmann trajectories on the Kadanoff-Baym contour, and $\boldabar=\{\boldabar_{\uparrow},\boldabar_{\downarrow}\},\bolda=\{\bolda_{\uparrow},\bolda_{\downarrow}\}$ for briefness. The measure is
\begin{align}
\mathcal{D}[\boldabar,\bolda]=\prod_{\sigma,t}\dd{\abar_{\sigma}(t)}\dd{a_{\sigma}(t)}e^{-\abar_{\sigma}(t)a_{\sigma}(t)}.
\end{align}
$\gK$ is only determined by $\Himp$, which can be formally written as
\begin{align}\label{eq:gKc}
\gK\left[\boldabar, \bolda \right] = e^{-\im\int_{\branch} dt \gHimp(t) }.
\end{align}
Here $\gHimp$ is obtained from $\Himp$ by making the substitutions $\aop_{\sigma}(t)\to a_{\sigma}(t)$, $\adop_{\sigma}(t)\to \abar_{\sigma}(t)$, $dt$ should be understood as branch-dependent, that is, $dt=\pm\delta t$ on $\branch^{\pm}$, and $dt=-\im\delta\tau$ on $\branch^{\circ}$, where $\delta t$ and $\delta \tau$ are the time step size on the real-time and imaginary-time axes respectively.
$\gI_{\sigma}$ is the Feynman-Vernon IF determined by $\Hint$ (more concretely, $J(\varepsilon)$), which can be formally written as 
\begin{align}\label{eq:gIc}
  \gI_{\sigma}[\boldabar_{\sigma},\bolda]=e^{-\int_{\branch}dt\int_{\branch}d t'\abar_{\sigma}(t)\Delta(t,t')a_{\sigma}(t')}.
\end{align}
$\Delta(t,t')$ is the hybridization function which encodes all the bath effects and can be calculated by
\begin{align}
  \label{eq:hybridization}
  \Delta(t,t')=\im\int\dd{\varepsilon}J(\varepsilon)D_{\varepsilon}(t,t').
\end{align}
with $D_{\varepsilon}(t,t')$ the free bath contour-ordered Green's function defined as
\begin{align}
  D_{\varepsilon}(t,t')\EqDef-\im\expval*{T_{\branch}\cop_{\varepsilon}(t)\cdop_{\varepsilon}(t')}_{\mathrm{bath}}.
\end{align}
Here $T_{\branch}$ is the contour-ordering operator that arranges operators on the contour in the order indicated by the arrows in Fig.~\ref{fig:kadanoff}, and $\expval{\cdots}_{\mathrm{bath}}$ means the expectation value with respect to the free bath.

\begin{figure*}[htbp]
  \centerline{\includegraphics[]{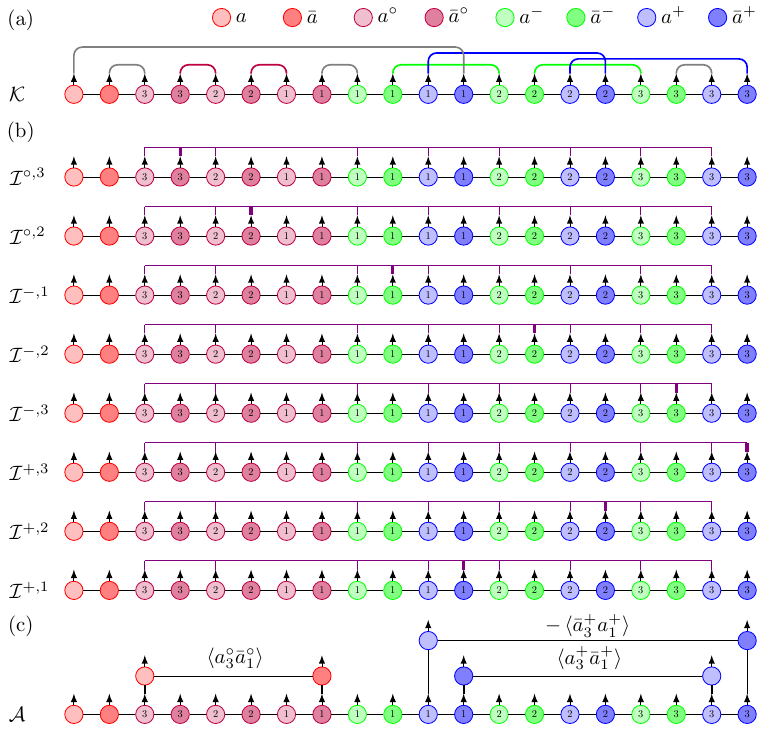}}
  \caption{Schematic illustration of the construction of (a) $\gK$ and (b) each partial influence functional as Grassmann MPSs for the non-interacting Toulouse model with $M=N=2$, where only a single spin needs to be considered and the spin indices are omitted. The purple, green, blue solid circles with numbers represent Grassmann variables at different time steps on the imaginary, backward and forward branches respectively. The pair of red solid circles at the left boundary without numbers represent the additional pair of GVs introduced to take care of the boundary condition in the impurity path integral. In panel (b) we have shown the concrete patterns of all the $M+2(N+1)=8$ partial IFs, which are multiplied together to obtain $\gI$ as a GMPS. (c) Calculating the Matsubara Green's function $\mathcal{G}_{3,1}$, the greater Green's function $G^>_{3,1}$ and the lesser Green's function $G^<_{3,1}$ based on the augmented density tensor $\gA$, corresponding to Eq.(\ref{eq:matsubara}), Eq.(\ref{eq:greater}) and Eq.(\ref{eq:lesser}) respectively.}
  \label{fig:demo}
\end{figure*}

\subsection{The mixed-time GTEMPO method}
\label{subsec:gtempo}

The continuous expressions in Eq.(\ref{eq:gKc}) and Eq.(\ref{eq:gIc}) are more of notational convenience. In actual calculation, the continuous integral along $\branch$ should be understood in the discrete sense where $\branch$ is first broken into small finite pieces, and then one sums over the contributions from all the possible discrete paths formed by these pieces. 
In the following, we introduce in detail the discretization scheme we have used in our numerical implementation, which is in the same spirit as our implementations of the real-time~\cite{ChenGuo2024a} and imaginary-time~\cite{ChenGuo2024b} GTEMPO methods. 

We denote $N=t/\delta t$, $M = \beta/\delta\tau$ as the total numbers of discrete time steps on the real- and imaginary-time axes respectively. We first discretize the continuous Grassmann trajectories into discrete ones as $\boldabar_{\sigma}^{\pm}=\{\abar_{\sigma,N+1}^{\pm},\cdots,\abar_{\sigma,1}^{\pm}\}$, $\bolda_{\sigma}^{\pm}=\{a_{\sigma,N+1}^{\pm},\cdots,a_{\sigma,1}^{\pm}\}$, $\boldabar_{\sigma}^{\circ}=\{\abar_{\sigma,M+1}^{\circ},\cdots,\abar_{\sigma,1}^{\circ}\}$ and $\bolda_{\sigma}^{\circ}=\{a_{\sigma,M+1}^{\circ},\cdots,a_{\sigma,1}^{\circ}\}$, where $a_{\sigma, j}^{\zeta}$ and $\abar_{\sigma, j}^{\zeta}$ are a pair of conjugate Grassmann variables (GVs) at time step $j$ on $\mathcal{C}^{\zeta}$.
We will also use $\boldabar_j^{\zeta}=\{\abar_{\uparrow,j}^{\zeta},\abar_{\downarrow,j}^{\zeta}\}$ and $\bolda_j^{\zeta}=\{a_{\uparrow,j}^{\zeta},a_{\downarrow,j}^{\zeta}\}$ to denote the set of GVs for both spins at the same time step $j$. In addition, we introduce a pair of GVs $a_{\sigma}$ and $\abar_{\sigma}$ to account for the boundary condition \cite{ChenGuo2024b}, which corresponds to the final trace in Eq.(\ref{eq:Zimp}). As a result, there will be $L = 8(N+1)+4(M+1)+4$ GVs in total for the single-orbital AIM, which will also be the size of all the GMPSs involved in the mixed-time GTEMPO calculations. 
In our numerical implementation, those GVs are ordered as
\begin{align}\label{eq:ordering}
&a_{\uparrow}\abar_{\uparrow}a_{\downarrow}\abar_{\downarrow}a^{\circ}_{\uparrow, M+1}\abar^{\circ}_{\uparrow, M+1}a^{\circ}_{\downarrow, M+1}\abar^{\circ}_{\downarrow, M+1}\cdots a^{\circ}_{\uparrow, 1}\abar^{\circ}_{\uparrow, 1}a^{\circ}_{\downarrow, 1}\abar^{\circ}_{\downarrow, 1} \nonumber \\ 
& a^{-}_{\uparrow, 1}\abar^{-}_{\uparrow, 1}a^{-}_{\downarrow, 1}\abar^{-}_{\downarrow, 1} a^{+}_{\uparrow, 1}\abar^{+}_{\uparrow, 1}a^{+}_{\downarrow, 1}\abar^{+}_{\downarrow, 1} \cdots a^{-}_{\uparrow, N+1}\abar^{-}_{\uparrow, N+1} \nonumber \\
& a^{-}_{\downarrow, N+1}\abar^{-}_{\downarrow, N+1} a^{+}_{\uparrow, N+1}\abar^{+}_{\uparrow, N+1}a^{+}_{\downarrow, N+1}\abar^{+}_{\downarrow, N+1} ,
\end{align}
where we have arranged the $\pm$ branches in nearly positions, different from the original time order in Fig.~\ref{fig:kadanoff}.

Based on the above notations, the discretized expression for $\gK$ can be written as
\begin{align}
  \label{eq:gK}
  \gK[\boldabar,\bolda]=&\braket{ -\bolda}{\bolda_{M+1}^{\circ}}\mel{\bolda_{M+1}^{\circ}}{\Uimp^{\circ}}{\bolda_M^{\circ}} \cdots\mel{\bolda_2^{\circ}}{\Uimp^{\circ}}{\bolda_1^{\circ}} \nonumber \\
  &\times \braket{\bolda_1^{\circ}}{a_1^-}\mel{\bolda_1^-}{\Uimp^-}{\bolda_2^-}\cdots\mel{\bolda_{N}^-}{\Uimp^-}{\bolda_{N+1}^-} \nonumber \\
  &\times\braket{\bolda_{N+1}^-}{\bolda_{N+1}^+}\mel{\bolda_{N+1}^+}{\Uimp^+}{\bolda_{N}^+}\cdots \nonumber \\
  &\times\mel{\bolda_2^+}{\Uimp^+}{\bolda_1^+}\braket{\bolda_1^+}{\bolda},
\end{align}
where $\bolda = \{a_{\uparrow}, a_{\downarrow}\}$, $\boldabar = \{\abar_{\uparrow}, \abar_{\downarrow}\}$ (the first and last term on the rhs take care of the boundary condition), and the discrete propagator $\Uimp^{\zeta} = e^{-\im \Himp dt}$ (noticing that $dt$ is branch-dependent). For single-orbital AIM, the propagators can be exactly evaluated as:
\begin{align}\label{eq:propagator}
\mel{\bm{\xi}}{\Uimp^{\zeta}}{\bm{\xi}'} &= e^{g^{\zeta}\sum_{\sigma}\bar{\xi}_{\sigma}\xi'_{\sigma}+(g^{\zeta})^2(e^{-\im dt U}-1)\bar{\xi}_{\uparrow}\bar{\xi}_{\downarrow}\xi'_{\downarrow}\xi'_{\uparrow}}
\end{align}
with $g^{\zeta}=e^{-\im dt(\varepsilon_d-U/2)}$ (See Appendix.~\ref{app:prop} for the derivation of this equation). 
With the ordering of GVs in Eq.(\ref{eq:ordering}), $\gK$ can be exactly built as a GMPS with bond dimension $16$.
For more sophisticated impurity models, one could also easily adapt the algorithm proposed in Ref.~\cite{ChenGuo2024b} to obtain a numerically exact expression for $\gK$.

The IF can be discretized using the quasi-adiabatic propagator path integral (QuAPI) method~\cite{makarov1994-path,makri1995-numerical}, which results in a discrete expression in the form
\begin{align}\label{eq:IF}
\gI_{\sigma} \approx e^{-\sum_{\zeta,\zeta'}\sum_{jk}\abar^{\zeta}_{\sigma,j}\Delta_{j,k}^{\zeta \zeta'}a^{\zeta'}_{\sigma,k}}.
\end{align}
There are $9$ hybridization matrices in total, and their expressions can be found in Appendix.~\ref{app:quapi}.

In this work, we build each $\gI_{\sigma}$ as a GMPS using the partial-IF algorithm as described in Refs.~\cite{ChenGuo2024a,GuoChen2024d}. Concretely, we first rewrite Eq.(\ref{eq:IF}) as
\begin{align}
  \label{eq:partial-IF}
\gI_{\sigma} \approx \prod_{\zeta, j} \gI_{\sigma}^{\zeta, j} \EqDef \prod_{\zeta,j} \left(e^{-\sum_{\zeta', k}\abar^{\zeta}_{\sigma,j}\Delta_{j,k}^{\zeta \zeta'}a^{\zeta'}_{\sigma,k}} \right),
\end{align}
then each partial-IF $\gI_{\sigma}^{\zeta, j}$ can be exactly written as a GMPS with bond dimension $2$ only~\cite{GuoChen2024d}. As a result, we can build $\gI_{\sigma}$ as a GMPS by multiplying $M+2(N+1)$ GMPSs, during which MPS bond truncation is performed to keep the bond dimension of the resulting GMPS to be within a given threshold $\chi$~\cite{ThoennissAbanin2023b,ChenGuo2024c}. The cost of this construction roughly scales as $O[(M+2N)^2\chi^3]$ as a general feature of the partial-IF algorithm~\cite{GuoChen2024d}.

The procedures to build $\gK$ and each partial IF as GMPSs are schematically shown in Fig.~\ref{fig:demo}(a,b) respectively, specialized for the noninteracting case such that we only need to consider a single spin for briefness. For the single-orbital AIM the construction of $\gK$ will contain $4$-body terms, while the partial IFs for the two spins have exactly the same patterns as in Fig.~\ref{fig:demo}(b), but acting on different spins separately. 

It should be noted that there is a stark difference between GTEMPO and the conventional wave-function based methods. In the wave-function based methods, the whole system evolves iteratively from real-time $t$ to $t+\delta t$ or from imaginary-time $\tau$ to $\tau+\delta\tau$, and the errors would generally accumulate with time. In contrast, in GTEMPO the whole time evolution window is addressed simultaneously, therefore in principle the errors will spread the whole time evolution window and do not necessarily grow with time. 
For the same reason, the errors generated on the real- and imaginary-time axes will generally affect each other, and the causality relation in the real-time axis will not be explicitly preserved.

\begin{figure}[htbp]
\begin{tikzpicture}[%
  inner sep=2mm,
    action/.style={rectangle, draw, fill=green!20, 
    text width=18em, rounded corners, align=justify},
  title/.style={font=\LARGE\scshape,node distance=16pt, text=black!40, inner sep=1mm}
]

  \node[action] (mapping) {Write down the analytical expression of the impurity partition function as a path integral of Grassmann trajectories, as in Eqs.(\ref{eq:PI}, \ref{eq:gKc}, \ref{eq:gIc}).};
  \node[action] (diagonalize) [below=of mapping] {Discretize $\gK$ into the product of discrete propagators as in Eq.(\ref{eq:gK}) and then build $\gK$ as a GMPS by multiplying these propagators onto the Grassmann vacuum. Each propagator is a Grassmann tensor of at most $4$ GVs, which can either be written down analytically as in Eq.(\ref{eq:propagator}) for the single-orbital AIM, or numerically using the algorithm in Ref.~\cite{ChenGuo2024b} in the general case.};
  \node[action] (reduce) [below=of diagonalize] {Discretize $\gI_{\sigma}$ using the QuAPI method (see Appendix.~\ref{app:quapi} for the final results of QuAPI), which gives Eq. \eqref{eq:IF}, and decompose it into the product of partial IFs as in Eq.(\ref{eq:partial-IF}). Build each partial-IF as a GMPS with bond dimension $2$ using the analytical expression in Ref.~\cite{GuoChen2024d}, and then multiply these partial-IFs together using GMPS multiplication~\cite{ChenGuo2024a} to obtain the GMPS representation of $\gI_{\sigma}$.};
  \node[action] (analytic) [below=of reduce] {Calculate multi-time impurity correlations based on the GMPS representations of $\gK$ and $\gI_{\sigma}$, for example, in Eqs.(\ref{eq:matsubara}, \ref{eq:greater}, \ref{eq:lesser}), where the ADT $\gA$ is calculated on the fly.};

  \draw[->] (mapping.south) -- (diagonalize);
  \draw[->] (diagonalize) -- (reduce);
  \draw[->] (reduce) -- (analytic);

  \node[fit=(mapping)(diagonalize)(reduce)(analytic)] (chart) {};

\end{tikzpicture}
\caption{The elementary procedures of the mixed-time GTEMPO method.}
\label{fig:illustration}
\end{figure}

\subsection{Calculating Green's functions}
\label{subsec:green-function}
Once $\gK$ and $\gI_{\sigma}$ are both built as GMPSs, their product gives the GMPS representation of the augmented density tensor (ADT):
\begin{align}
  \label{eq:ADT}
\gA[\boldabar,\bolda]=\gK[\boldabar,\bolda]\prod_{\sigma}\gI_{\sigma}[\boldabar_{\sigma},\bolda_{\sigma}],
\end{align}
which is the integrand in Eq.(\ref{eq:PI}). 
The ADT contains the information of the impurity dynamics on the whole Kadanoff-Baym contour, based on which
any multi-time 
impurity correlation functions can be calculated straightforwardly following the standard textbook definitions~\cite{negele1998-quantum}. For
example, the Matsubara Green's function $\mathcal{G}_{j, k}$  between two imaginary-time steps $j$, $k$ ($j > k$) can be calculated as
\begin{align}\label{eq:matsubara}
-\mathcal{G}_{j,k} &\EqDef \expval*{\aop_{\sigma, j}\adop_{\sigma, k}} = \frac{\int \mea [\boldabar, \bolda] a_{\sigma, j}^{\circ} \abar_{\sigma, k}^{\circ} \gA[\boldabar, \bolda]}{\Zimp},
\end{align}
the greater Green's function $G^>_{j,k}$ and lesser Green's function $G^<_{j,k}$ can be calculated as 
\begin{align}
\im G^{>}_{j,k} &\EqDef \expval*{\aop_{\sigma, j}\adop_{\sigma, k}} = \frac{\int \mea [\boldabar, \bolda] a_{\sigma, j}^+ \abar_{\sigma, k}^+ \gA[\boldabar, \bolda]}{\Zimp} ; \label{eq:greater} \\
-\im G^{<}_{j,k} &\EqDef \expval*{\adop_{\sigma, k} \aop_{\sigma, j}} = \frac{\int \mea [\boldabar, \bolda] \abar_{\sigma, k}^- a_{\sigma, j}^+ \gA[\boldabar, \bolda]}{\Zimp}. \label{eq:lesser}
\end{align}
With $G^>_{j,k}$ and $G^<_{j,k}$ , the retarded Green's function $G_{j,k}$ can also be easily obtained from its definition
\begin{align}\label{eq:retarded}
G_{j, k}\EqDef G^{>}_{j,k} - G^{<}_{j,k}.
\end{align}
It should be noted here that in practice $\gA$ is not built directly, but computed on the fly using a zipup algorithm for efficiency~\cite{ChenGuo2024a,ChenGuo2024b}.

The elementary procedures of the mixed-time GTEMPO method is summarized in Fig.~\ref{fig:illustration}.

\section{Numerical results}

In the following we demonstrate the performance of the mixed-time GTEMPO method in the noninteracting case and in the single-orbital AIM.
In our numerical simulations for this work we will focus on a semi-circular bath spectrum density 
\begin{align}
J(\varepsilon)=\frac{\Gamma D}{2\pi}\sqrt{1-(\varepsilon/D)^2},
\end{align}
with $D=2$ and $\Gamma=0.1$, and we will take $\Gamma$ as the unit.

\subsection{The noninteracting ($U=0$) limit}

We first validate the mixed-time GTEMPO method on the equilibrium Green's functions in the noninteracting Toulouse model~\cite{LeggettZwerger1987,mahan2000-many}, where we focus on the half filling case with $\varepsilon_d = 0$.

\begin{figure}
  \includegraphics[width=\columnwidth]{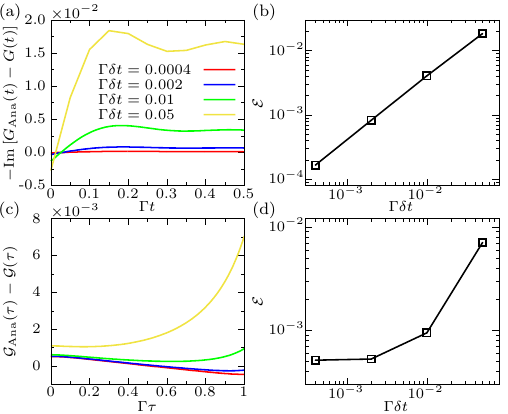} 
  \caption{(a, c) The deviation of the retarded Green's function from the analytical solution $G_{\rm Ana}(t)$ as a function of $t$ in (a) and the deviation of the Matsubara Green's function from the analytical solution $\mathcal{G}_{\rm Ana}(\tau)$ as a function of $\tau$ in (c) for the noninteracting Toulouse model, where we have fixed $\Gamma\delta \tau=0.01$. The red, blue, green and yellow lines correspond to the mixed-time GTEMPO calculations with $\Gamma\delta t=0.0004, 0.002, 0.01, 0.05$ respectively. (b, d) Scaling of the maximum errors of the retarded (b) and the Matsubara (d) Green's functions against $\delta t$.
  We have used $\chi=120$ in these mixed-time GTEMPO simulations.}
    \label{fig:fig1a}
\end{figure}

\begin{figure}
  \includegraphics[width=\columnwidth]{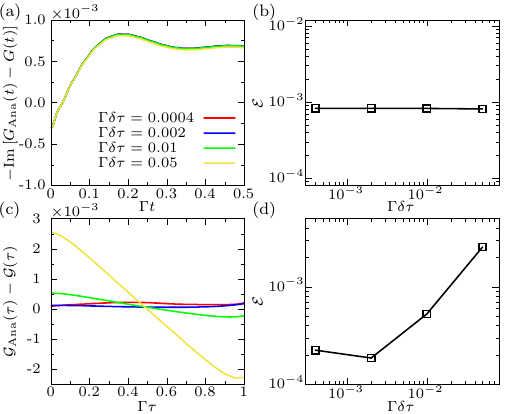} 
  \caption{(a, c) The deviation of the retarded Green's function from the analytical solution as a function of $t$ in (a) and the deviation of the Matsubara Green's function from the analytical solution as a function of $\tau$ in (c), where we have fixed $\Gamma \delta t=0.002$. The red, blue, green and yellow lines correspond to the mixed-time GTEMPO calculations with $\Gamma\delta \tau=0.0004, 0.002, 0.01, 0.05$ respectively. (b, d) Scaling of the maximum errors of the retarded (b) and the Matsubara (d) Green's functions against $\delta \tau$. We have used $\chi=120$ in these mixed-time GTEMPO simulations.}
    
    \label{fig:fig1b}
\end{figure}

In Fig.~\ref{fig:fig1a} and Fig.~\ref{fig:fig1b}, we study the accuracies of the retarded and the Matsubara Green's functions, compared to the analytical solutions (See Appendix.~\ref{app:ana} for the analytical expressions of these Green's functions in the noninteracting limit), against the discrete time step sizes $\delta t$ and $\delta\tau$ respectively. 
The solid lines in both figures are the errors between mixed-time GTEMPO results and the corresponding analytic solutions. For both figures we have used $\Gamma t_f=0.5$ and $\Gamma \beta=1$, we have also used a large bond dimension $\chi=120$ such that the MPS bond truncation error can be essentially ignored (we will see in the later simulations that we can already obtain very accurate results with $\chi \leq 120$ for even larger $\beta$).
In Fig.~\ref{fig:fig1a}, we fix $\Gamma\delta \tau=0.01$ and tune $\Gamma\delta t=0.0004$ (the red lines), $\Gamma\delta t=0.002$ (the blue lines), $\Gamma\delta t=0.01$ (the green lines), $\Gamma\delta t=0.05$ (the yellow lines). In panel (a), we plot the errors of the retarded Green's function as a function of the real time $t$, and in panel (c) we plot the errors of the Matsubara Green's function as a function of the imaginary time $\tau$, compared to their analytical solutions. We can see that both the retarded and the Matsubara Green's functions calculated by mixed-time GTEMPO become more accurate with smaller $\delta t$. In panels (b, d), we show the maximum errors between the mixed-time GTEMPO results and the analytical solutions as functions of $\delta t$, corresponding to panels (a,c) respectively. The maximum error is defined as
\begin{align}
  \mathcal{E}(\vec{x},\vec{y})=\mathrm{max}_i|x_i-y_i|
\end{align}
for $x_i$ and $y_i$ being the elements of the two vectors $\vec{x}$ and $\vec{y}$. 
From Fig.~\ref{fig:fig1a}(b), we can see that the maximum error of the retarded Green's function scales linearly against $\delta t$. 
This indicates that the accuracy of $G(t)$ is dominated by the error generated by $\delta t$ under the given set of parameters. 
Mathematically, the accuracy of the Matsubara Green's function $\mathcal{G}(\tau)$ should be independent of $\delta t$. However, as mentioned earlier, this causality relation is not explicitly preserved in GTEMPO, and thus the accuracy of $\mathcal{G}(\tau)$ can also be improved by smaller $\delta t$ as can be seen from Fig.~\ref{fig:fig1a}(d). This improvement, as expected, is not linear since the accuracy of $G(\tau)$ is also significantly affected by finite $\delta\tau$.

In Fig.~\ref{fig:fig1b}, we fix $\Gamma \delta t=0.002$, and tune $\Gamma \delta \tau=0.0004$ (the red lines), $\Gamma \delta \tau=0.002$ (the blue lines), $\Gamma\delta \tau=0.01$ (the green lines), and $\Gamma\delta \tau=0.05$ (the yellow lines). Similar to Fig.~\ref{fig:fig1a}, we plot the errors of the retarded and the Matsubara Green's functions compared to their analytical solutions in panels (a, c) respectively, and plot the scaling of the corresponding maximum errors against $\delta \tau$ in panels (b, d). 
From Fig.~\ref{fig:fig1b}(b), we can see that the maximum error of the retarded Green's function is almost unaffected by $\delta\tau$.
From Fig.~\ref{fig:fig1b}(d), we can see that the maximum error of the Matsubara Green's function decreases with smaller $\delta\tau$, but not linearly, similar to Fig.~\ref{fig:fig1a}(d), which shows that both $\delta t$ and $\delta \tau$ significantly affect the accuracy of $\mathcal{G}(\tau)$ under the given set of parameters.
There is a small increase of $\mathcal{G}(\tau)$ when $\Gamma\delta \tau$ decreases from $0.002$ to $0.0004$, which is likely a coincidence due to the interplay of the simultaneous influences of finite $\delta t$ and $\delta \tau$ when their effects are comparable. To this end, we would like to point out that since there are two time discretizations, $\delta t$ and $\delta \tau$, and the errors on the real- and imaginary-time axes will generally affect each other in our algorithm, we can not expect linear scaling against one of the time discretization unless we make the other to be extremely small (which, however, may cause numerical instabilities in the MPS arithmetics).

\begin{figure*}
  \includegraphics[width=\textwidth]{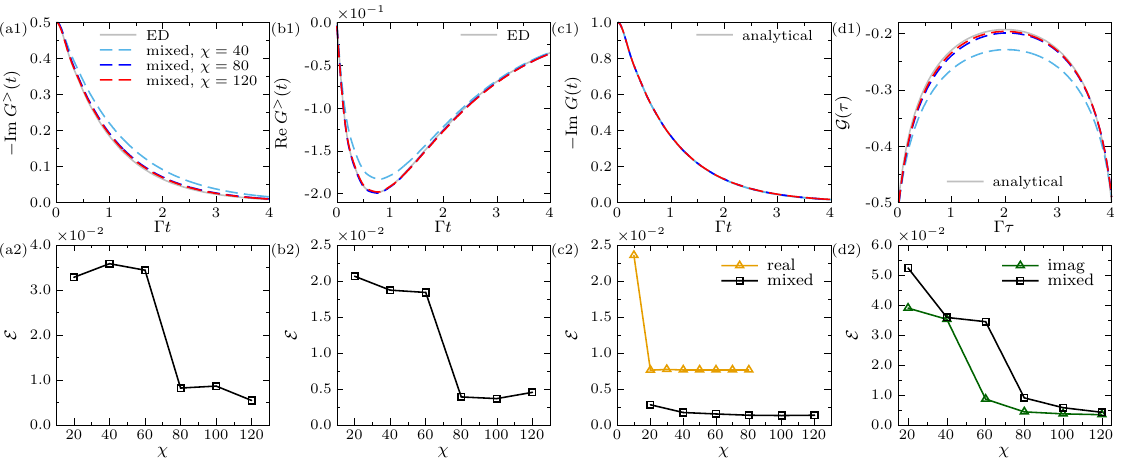} 
  \caption{(a1, b1, c1, d1) The imaginary (a1) and real (b1) parts of the greater Green's function $G^>(t)$, the imaginary part of the retarded Green's function $G(t)$ (c1), and the Matsubara Green's function $\mathcal{G}(\tau)$ (d1) of the Toulouse model as functions of the real (a1,b1,c1) or imaginary (d1) time. The cyan, blue and red dashed lines are mixed-time GTEMPO results with bond dimensions $\chi=40,80,120$ respectively. The gray solid lines are the ED results in (a1, b1) and the analytical solutions in (c1, d1). (a2, b2, c2, d2) The black solid line with square show the maximum errors of between the mixed-time GTEMPO results and the corresponding exact solutions in (a1,b1,c1,d1) as functions of the bond dimension $\chi$. The yellow solid line with triangle in panel (c2) shows the maximum error between the real-time GTEMPO results and the analytical solutions, and the green solid line with triangle in panel (d2) shows the maximum error between the imaginary-time GTEMPO results and the analytical solutions. In these simulations we have used $\Gamma\delta t=0.005$ and $\Gamma\delta \tau=0.01$ for mixed-time GTEMPO, $\Gamma\delta t=0.005$ for real-time GTEMPO, and $\Gamma\delta \tau=0.01$ for imaginary-time GTEMPO.
    }
    \label{fig:fig2}
\end{figure*}

In the following we will fix $\Gamma\delta t=0.005$ and $\Gamma\delta \tau=0.01$ for mixed-time GTEMPO simulations. We will also set $\Gamma\beta=4$.
In Fig.~\ref{fig:fig2}, we study the accuracy of the greater, the lesser, the retarded and the Matsubara Green's functions against the bond dimension $\chi$.
In Fig.~\ref{fig:fig2}(a1, b1), we benchmark the imaginary and real parts of $G^>(t)$ against the exact diagonalization (ED) results. $G^<(t)$ is not shown as it has the same real part as and the opposite imaginary part to $G^>(t)$ at half filling. In Fig.~\ref{fig:fig2}(c1, d1), we benchmark the imaginary part of $G(t)$ (the real part vanishes at half filling and is not shown), and $\mathcal{G}(\tau)$, against the corresponding analytical solutions respectively. The dashed lines in Fig.~\ref{fig:fig2}(a1, b1, c1, d1) are mixed-time GTEMPO results calculated using different bond dimensions $\chi=40,80,120$, while the gray solid lines are the ED or analytical results. 
For ED we have discretized the bath into $2000$ equal-distant frequencies ($\delta\varepsilon/\Gamma = 0.005$) and we have verified that the ED results have well converged against bath discretization (which is the only error in ED). In fact with such a bath discretization, the absolute difference between the ED results and the analytical solutions for the retarded Green's function is smaller than $10^{-6}$. In Fig.~\ref{fig:fig2}(a2, b2, c2, d2), we show the maximum errors between the mixed-time GTEMPO results and corresponding exact solutions (e.g., ED results in (a1, b1) and analytical solutions in (c1, d1)) as a function of $\chi$.
We can see that with $\chi=80$ both $G^>(t)$ and $\mathcal{G}(\tau)$ calculated by mixed-time GTEMPO already converge quite well with ED, with the maximum error within $1\%$. 
From Fig.~\ref{fig:fig2}(c1, c2), we can see that the retarded Green's function, calculated from Eq.(\ref{eq:retarded}), converges much faster than $G^>(t)$, which means that the errors in $G^>(t)$ and $G^<(t)$ perfectly cancel each other even with a very small bond dimension $\chi=20$,
this is likely due to a coincidence that the noninteracting retarded Green's function is independent of the initial state~\cite{mahan2000-many}. 
The dependence of the maximum error in the Matsubara Green's function on $\chi$ is comparable to the imaginary-time GTEMPO calculations. Crucially, even though in the mixed-time GTEMPO calculations we deal with more GVs compared to the real- and imaginary-time GTEMPO calculations, we observe that the bond dimension required to obtain converged and accurate results seem to be not larger than the latter ones, thus the mixed-time GTEMPO calculation will not be a lot more expensive than the imaginary-time calculation (it has been observed that the imaginary-time GTEMPO calculations usually require larger bond dimensions than the real-time GTEMPO calculations to achieve similar accuracy~\cite{ChenGuo2024c}), as the computational cost of the partial-IF algorithm roughly scales quadratically with the number of GVs, and cubically with $\chi$~\cite{GuoChen2024d}.

\subsection{The single-orbital Anderson impurity model}

\begin{figure}
  \includegraphics[width=\columnwidth]{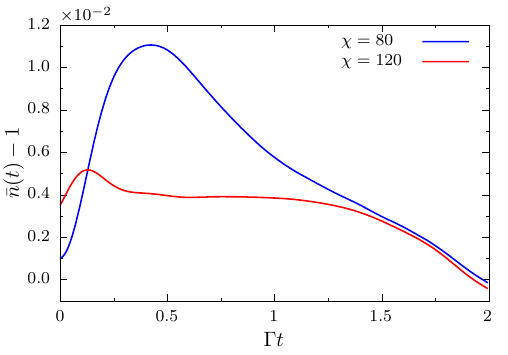} 
  \caption{The average occupation $\bar{n}$ as a function of real time $t$ for the single-orbital Anderson impurity model for $\chi=80$ (the blue line) and $\chi=120$ (the red line).}
    \label{fig:density}
\end{figure}

\begin{figure*}
  \includegraphics[width=\textwidth]{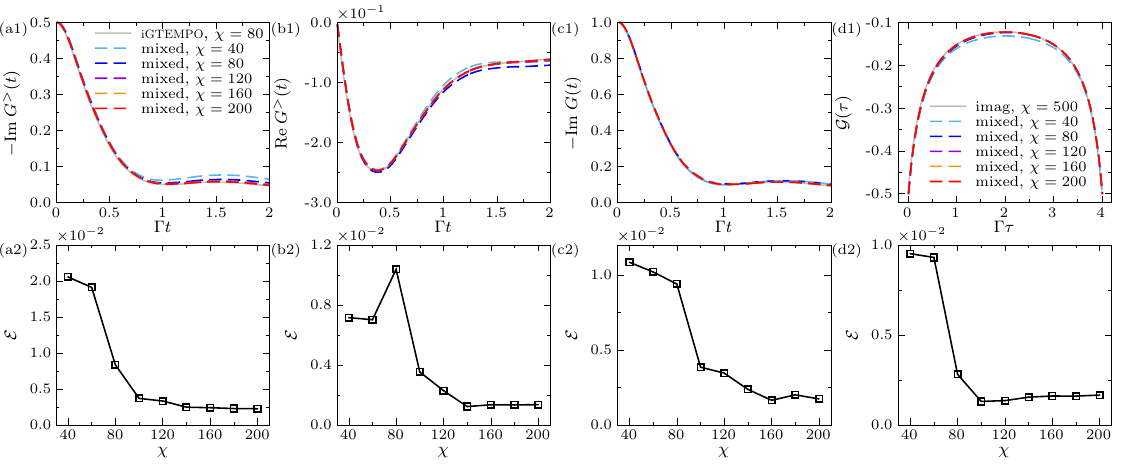} 
  \caption{(a1, b1, c1, d1) The  imaginary (a1) and real (b1) parts of the greater Green's function $G^>(t)$, the imaginary part of the retarded Green's function $G(t)$ (c1), and the Matsubara Green's function $\mathcal{G}(\tau)$ (d1) of the single-orbital Anderson impurity model as functions of the real (a1,b1,c1) or imaginary (d1) time. The gray solid lines in (a1,b1,c1) are the infinite-time GTEMPO results calculated using $\chi=80$ and $\Gamma\delta t =0.005$. The gray solid line in (d1) is the imaginary-time GTEMPO results calculated with $\chi=500$ and $\Gamma\delta \tau =0.01$.
  (a2, b2, c2) The black solid line with square show the maximum errors between the mixed-time GTEMPO results and the baseline infinite-time GTEMPO results in (a1,b1,c1) against $\chi$. (d2) The black solid line with square show the maximum errors between the mixed-time GTEMPO results and the baseline imaginary-time GTEMPO results in (d1) against $\chi$.
  In these mixed-time GTEMPO calculations we have used $U/\Gamma=5$, $\Gamma\delta t=0.005$ and $\Gamma\delta \tau=0.01$.
    }
    \label{fig:fig3}
\end{figure*}

Next we proceed to study the single-orbital AIM, and we still focus on the half-filling case with $\varepsilon_d=0$ and $U/\Gamma = 5$. 

As a first check of the quality of the equilibrium state, we plot the average occupation $\bar{n}(t)\EqDef\expval*{\adop_{\sigma}(t) \aop_{\sigma}(t)}$ 
as a function of the real time $t$ in Fig.~\ref{fig:density} for $\chi=80, 120$. Ideally, $\bar{n}(t)$ should be exactly equal to $1$ as we have chosen the half filling condition. We can see that the maximum derivation of $\bar{n}$ from $1$ is around $1\%$, which is approximately the same order as $\delta\tau$, and that the maximum derivation decreases by half when increasing $\chi$ from $80$ to $120$.

For the single-orbital AIM, there does not exist exact solutions for the Green's functions. Nevertheless, for the Matsubara Green's function we can benchmark the accuracy against imaginary-time GTEMPO calculations, and for real-time Green's functions we can benchmark the accuracy against the infinite-time GTEMPO (referred to as iGTEMPO) calculations in Ref.~\cite{GuoChen2024e} which directly aims at the steady state. 
Similar to Fig.~\ref{fig:fig2}, we show the imaginary and real parts of $G^>(t)$, the imaginary part of $G(t)$, and $\mathcal{G}(\tau)$ in Fig.~\ref{fig:fig3}(a1, b1, c1, d1) respectively. The gray solid lines in Fig.~\ref{fig:fig3}(a1,b1,c1) are the infinite-time GTEMPO results calculated with $\chi=80$ and $\Gamma\delta t =0.005$, which are also used as the baseline to calculate the maximum errors in Fig.~\ref{fig:fig3}(a2,b2,c2). The gray solid line in Fig.~\ref{fig:fig3}(d1) shows the imaginary-time GTEMPO results calculated with $\chi=500$ and $\Gamma\delta \tau =0.01$, which is also used as the baseline to calculate the maximum errors in Fig.~\ref{fig:fig3}(d2).
Again we can see that $G^>(t)$ calculated by mixed-time GTEMPO using $\chi=80$ already agrees well with the iGTEMPO calculations, with maximum error within $1\%$. Meanwhile, the maximum error in $G^>(t)$ keeps decreasing with $\chi$, and only approximately converges at $\chi=200$. For $\mathcal{G}(\tau)$, the results calculated by mixed-time GTEMPO agree very well with the imaginary-time GTEMPO calculations at $\chi=80$ (the maximum error is less than $0.3\%$, compared to $1\%$ in the noninteracting case), which agrees with the previous observation that the imaginary-time GTEMPO calculations become more accurate for larger $U$~\cite{ChenGuo2024b}. 
From Fig.~\ref{fig:fig3}(b2, c2), we can see that the errors of the retarded Green's function against $\chi$ in the interacting case is similar to $G^>(t)$ (which means that the errors in $G^>(t)$ and $G^<(t)$ no longer perfectly cancel each other), in contrast with the noninteracting case.

\section{Discussions}
In summary, we have proposed a mixed-time GTEMPO method that works on the L-shaped Kadanoff-Baym contour. The method complements the previous real-time GTEMPO and imaginary-time GTEMPO methods, which allows to directly consider correlated impurity-bath initial state. It contains only three hyperparameters, the real-time step size $\delta t$ and imaginary-time step size $\delta\tau$, plus the MPS bond dimension $\chi$.
In our numerical examples, we benchmark this method against exact solutions in the noninteracting case, and against the iGTEMPO and imaginary-time GTEMPO calculations in the single-orbital Anderson impurity model. 
We also show that the numerical errors of this method can be well suppressed by refining those three hyperparameters.
Crucially, we observe that the bond dimension required in the mixed-time GTEMPO calculations is not larger than that required in the imaginary-time calculations, even though the size of the GMPS and the number of GMPS multiplications become larger, therefore the computational overhead of the mixed-time GTEMPO method will not be significant compared to the imaginary-time GTEMPO method. Compared to the iGTEMPO method, the mixed-time GTEMPO method will be less efficient, but it contains less hyperparameters than the latter.
This method could be a perfect benchmarking baseline for those alternatives which are more efficient but with less-controlled approximations. It can also be used as an impurity solver in DMFT, where one can obtain the real- and imaginary-time Green's functions simultaneously (see Appendix.~\ref{app:dmft} for the elementary steps of DMFT loop on the real-time axis based on the mixed-time GTEMPO).

\begin{acknowledgments}
This work is supported by National Natural Science Foundation of China under Grant No. 12104328. C. G. is supported by the Open Research Fund from State Key Laboratory of High Performance Computing of China (Grant No. 202201-00).
\end{acknowledgments}

\appendix

\section{Quasi-adiabatic propagator path integral on the Kadanoff-Baym contour}\label{app:quapi}
In this article we adopt the first-order QuAPI discretization scheme~\cite{makarov1994-path,makri1995-numerical}, and here we list the explicit expressions of the $9$ hybridization matrices used in our numerical simulations. We use $n(\varepsilon)=(e^{\beta\varepsilon}+1)^{-1}$ to denoted the Fermi-Dirac distribution function.
\begin{widetext}
\begin{equation}
  \Delta_{j,k}^{++}=\begin{cases}
    \displaystyle 2\int d\varepsilon \frac{J(\varepsilon)}{\varepsilon^2}[1-n(\varepsilon)]e^{-\im\varepsilon(j-k)\delta t}(1-\cos\varepsilon\delta t), & j > k;\\
    \displaystyle -2\int d\varepsilon \frac{J(\varepsilon)}{\varepsilon^2}n(\varepsilon)e^{-\im\varepsilon(j-k)\delta t}(1-\cos\varepsilon\delta t), & j < k;\\
    \displaystyle \int d\varepsilon \frac{J(\varepsilon)}{\varepsilon^2}
    \qty{[1-n(\varepsilon)][(1-\im\varepsilon\delta t)-e^{-\im\varepsilon\delta t}]
    -n(\varepsilon)[(1+\im\varepsilon\delta t)-e^{\im\varepsilon\delta t}]}, & j=k,
  \end{cases}
\end{equation}
\begin{equation}
  \Delta^{+-}_{j,k}=
    \displaystyle2\int d\varepsilon \frac{J(\varepsilon)}{\varepsilon^2}
    n(\varepsilon)e^{-\im\varepsilon(j-k)\delta t}(1-\cos\varepsilon\delta t),
\end{equation}
\begin{equation}
  \Delta^{+\circ}_{j,k}=-\int d\varepsilon \frac{J(\varepsilon)}{\varepsilon^2}
  n(\varepsilon)e^{-\im\varepsilon j\delta t}e^{\varepsilon k\delta\tau}
  (e^{-\im\varepsilon\delta t}-1)(e^{\varepsilon\delta\tau}-1),
\end{equation}
\begin{equation}
  \Delta^{-+}_{j,k}=\displaystyle -2\int d\varepsilon \frac{J(\varepsilon)}{\varepsilon^2}
    [1-n(\varepsilon)]e^{-\im\varepsilon(j-k)\delta t}(1-\cos\varepsilon\delta t),
\end{equation}
\begin{equation}
  \Delta^{--}_{j,k}=\begin{cases}
    \displaystyle-2\int d\varepsilon \frac{J(\varepsilon)}{\varepsilon^2}
    n(\varepsilon)e^{-\im\varepsilon(j-k)\delta t}(1-\cos\varepsilon\delta t), & j>k;\\
    \displaystyle 2\int d\varepsilon \frac{J(\varepsilon)}{\varepsilon^2}
    [1-n(\varepsilon)]e^{-\im\varepsilon(j-k)\delta t}(1-\cos\varepsilon\delta t), & j<k;\\
    \displaystyle-\int d\varepsilon \frac{J(\varepsilon)}{\varepsilon^2}
    \qty{n(\varepsilon)[(1-\im\varepsilon\delta t)-e^{-\im\varepsilon\delta t}]
    -[1-n(\varepsilon)][(1+\im\varepsilon\delta t)-e^{\im\varepsilon\delta t}]}, &j=k,
  \end{cases}
\end{equation}
\begin{equation}
  \Delta^{-\circ}_{j,k}=\int d\varepsilon \frac{J(\varepsilon)}{\varepsilon^2}
  n(\varepsilon)e^{-\im\varepsilon j\delta t}e^{\varepsilon k\delta\tau}(e^{-\im\varepsilon\delta t}-1)(e^{\varepsilon\delta\tau}-1),
\end{equation}
\begin{equation}
  \Delta^{\circ+}_{j,k}=\int d\varepsilon \frac{J(\varepsilon)}{\varepsilon^2}
  [1-n(\varepsilon)]e^{-\varepsilon j\delta\tau}e^{\im\varepsilon k\delta t}
  (e^{\im\varepsilon \delta t}-1)(e^{-\varepsilon \delta\tau}-1),
\end{equation}
\begin{equation}
  \Delta^{\circ-}_{j,k}=-\int d\varepsilon \frac{J(\varepsilon)}{\varepsilon^2}
  [1-n(\varepsilon)]e^{-\varepsilon j\delta\tau}e^{\im\varepsilon k\delta t}
  (e^{\im\varepsilon \delta t}-1)(e^{-\varepsilon \delta\tau}-1),
\end{equation}
\begin{equation}
  \Delta^{\circ\circ}_{j,k}=\begin{cases}
    \displaystyle-2\int d\varepsilon \frac{J(\varepsilon)}{\varepsilon^2}
    [1-n(\varepsilon)]e^{-\varepsilon(j-k)\delta\tau}(1-\cosh\varepsilon\delta\tau),&j>k;\\
    \displaystyle2\int d\varepsilon \frac{J(\varepsilon)}{\varepsilon^2}
    n(\varepsilon)e^{-\varepsilon(j-k)\delta\tau}(1-\cosh\varepsilon\delta\tau),&j<k;\\
    \displaystyle-\int d\varepsilon \frac{J(\varepsilon)}{\varepsilon^2}
    \qty{[1-n(\varepsilon)][(1-\varepsilon\delta\tau)-e^{-\varepsilon\delta\tau}]
    -n(\varepsilon)[(1+\varepsilon\delta\tau)-e^{\varepsilon\delta\tau}]}, & j=k.
  \end{cases}
\end{equation}
\end{widetext}

\section{Derivation of the discrete propagator for the single-orbital Anderson impurity model}\label{app:prop}
Here we give the derivation of the propagator \eqref{eq:propagator} of the single-orbital AIM. We denote the impurity Hamiltonian as
\begin{align}
    \Himp=&\Hop_0+\Hop_1 =(\varepsilon_d-\frac{1}{2}U)\sum_{\sigma}\adop_{\sigma}\aop_{\sigma}
    +U\adop_{\uparrow}\adop_{\downarrow}\aop_{\downarrow}\aop_{\uparrow}.
\end{align}
It is clear that $\Hop_0$ and $\Hop_1$ commute with each other,
and thus the discrete evolutionary operator can be written as
\begin{equation}
\Uimp^{\zeta}=e^{-\im\Himp dt}=e^{-\im\Hop_0dt}e^{-\im\Hop_1dt}.
\end{equation}
The propagator \eqref{eq:propagator} is then
\begin{equation}
  \label{eq:propagator-1}
  \mel{\bm{\xi}}{\Uimp^{\zeta}}{\bm{\xi}'}=\int\mea[\bm{\bar{\xi}}''\bm{\xi}'']
  \mel{\bm{\xi}}{e^{-\im\Hop_0dt}}{\bm{\xi}''}\mel{\bm{\xi}''}{e^{-\im\Hop_1dt}}{\bm{\xi}'},
\end{equation}
where the measure is
\begin{equation}
\mea[\bm{\bar{\xi}}''\bm{\xi}'']=\prod_{\sigma}d\bar{\xi}''_{\sigma}d\xi''_{\sigma}e^{-\bar{\xi}''_{\sigma}\xi''_{\sigma}}.
\end{equation}
According to the general properties of the Grassmann coherent state \cite{negele1998-quantum}
\begin{equation}
  \bra{\bm{\xi}}\adop_{\sigma}=\bra{\bm{\xi}}\bar{\xi}_{\sigma},\quad
  \aop_{\sigma}\ket{\bm{\xi}}=\xi_{\sigma}\ket{\bm{\xi}},
\end{equation}
and
\begin{equation}
  \braket{\bm{\xi}}{\bm{\xi}'}=e^{\sum_{\sigma}\bar{\xi}_{\sigma}\xi_{\sigma}}=
  (1+\bar{\xi}_{\uparrow}\xi_{\uparrow})(1+\bar{\xi}_{\downarrow}\xi_{\downarrow}),
\end{equation}
we have
\begin{equation}
  \begin{split}
  \mel{\bm{\xi}}{e^{-\im\Hop_0dt}}{\bm{\xi}''}=&
    \mel{\bm{\xi}}{\sum_{k=0}^{\infty}\frac{(-\im\Hop_0dt)^k}{k!}}{\bm{\xi}''}\\
    =&e^{g^{\zeta}\sum_{\sigma}\bar{\xi}_{\sigma}\xi''_{\sigma}},
  \end{split}
\end{equation}
and
\begin{equation}
  \begin{split}
    \mel{\bm{\xi}''}{e^{-\im\Hop_1dt}}{\bm{\xi}'}
    =&\mel{\bm{\xi}''}{\sum_{k=0}^{\infty}\frac{(-\im\Hop_1dt)^k}{k!}}{\bm{\xi}'}\\
    =&e^{\sum_{\sigma}\bar{\xi}''_{\sigma}\xi'_{\sigma}}
       e^{(e^{-\im dtU}-1)\bar{\xi}''_{\uparrow}\bar{\xi}''_{\downarrow}\xi'_{\downarrow}\xi'_{\uparrow}},
  \end{split}
\end{equation}
where $g^{\zeta}=e^{-\im dt(\varepsilon_d-U/2)}$ and we have used the fact that $(\adop_{\sigma}\aop_{\sigma})^k=\adop_{\sigma}\aop_{\sigma}$ for $k\ge1$. Therefore Eq. \eqref{eq:propagator-1} becomes
\begin{equation}
  \begin{split}
    \mel{\bm{\xi}}{\Uimp^{\zeta}}{\bm{\xi}'}
    =&\int\mea[\bm{\bar{\xi}}''\bm{\xi}'']e^{g^{\zeta}\sum_{\sigma}\bar{\xi}_{\sigma}\xi''_{\sigma}}e^{\sum_{\sigma}\bar{\xi}''_{\sigma}\xi'_{\sigma}}\\
     &\times e^{(e^{-\im dtU}-1)\bar{\xi}''_{\uparrow}\bar{\xi}''_{\downarrow}\xi'_{\downarrow}\xi'_{\uparrow}}\\
    =&\int\mea[\bm{\bar{\xi}}''\bm{\xi}''](1+g^{\zeta}\bar{\xi}_{\uparrow}\xi''_{\uparrow})
       (1+g^{\zeta}\bar{\xi}_{\downarrow}\xi''_{\downarrow})\\
    &\times(1+\bar{\xi}''_{\uparrow}\xi'_{\uparrow})(1+\bar{\xi}''_{\downarrow}\xi'_{\downarrow})\\
    &\times[1+(e^{-\im dtU}-1)\bar{\xi}''_{\uparrow}\bar{\xi}''_{\downarrow}\xi'_{\downarrow}\xi'_{\uparrow}],
  \end{split}
\end{equation}
expanding this expression and integrating out the GVs
$\bm{\bar{\xi}}'',\bm{\xi}''$ yields the desire formula \eqref{eq:propagator}.

\section{Analytic solutions in the noninteracting case}\label{app:ana}
In the noninteracting limit, the Coulomb interaction term is absent in the Hamiltonian and the Hamiltonian can then be written as
\begin{equation}
  \Hop=-\mu\adop\aop+\sum_k\varepsilon_k\cdop_k\cop_k+\sum_kV_k(\adop\cop_k+\cdop_k\aop),
\end{equation}
where $\mu$ is the chemical potential, and in the main text we have set $\mu=-(\varepsilon_d-U/2)$.

We denote the Matsubara Green's function in the noninteracting limit as $\mathcal{G}_0(\tau)$ whose definition is
\begin{equation}
\mathcal{G}_0(\tau)=-\expval*{T_{\tau}\aop(\tau)\adop}=-\frac{1}{Z}\Tr[e^{-\beta\Hop}T_{\tau}\aop(\tau)\adop],
\end{equation}
where $\aop(\tau)=e^{\tau\Hop}\aop e^{-\tau\Hop}$, $Z=\Tr e^{-\beta\Hop}$ is the partition function and $T_{\tau}$ is the imaginary-time ordering operator. It can be solved analytically in imaginary-frequency as \cite{mahan2000-many}
\begin{equation}
    \mathcal{G}_0^{-1}(\im\omega_s)=\im\omega_s+\mu-\Delta(\im\omega_s),
\end{equation}
where
\begin{equation}
  \Delta(\im\omega_s)=\int \frac{J(\varepsilon)}{\im\omega_s-\varepsilon}d\varepsilon
\end{equation}
is the hybridization function and
\begin{equation}
\omega_s=\frac{(2s+1)\pi}{\beta},\quad s=0,\pm1,\pm2,\cdots
\end{equation}
is the the Matsubara frequency for fermionic system.


The Matsubara Green's function has the asymptotic behavior that $\mathcal{G}_0(\im\omega_s)=\frac{1}{\im\omega_s}$ as $s\to\pm\infty$. Therefore we can evaluate it numerically by a truncation in Matsubara frequency with an asymptotic correction. That is, we replace $\mathcal{G}_0(\im\omega_s)$ out of the range $s_{\mathrm{min}}\le s\le s_{\mathrm{max}}$ by its asymptotic form for which
\begin{equation}
  \begin{split}
  \mathcal{G}_0(\tau)=\frac{1}{\beta}\sum_{s=s_{\mathrm{min}}}^{s_{\mathrm{max}}}
    \qty[\mathcal{G}_0(\im\omega_s)-\frac{1}{\im\omega_s}]e^{-\im\omega_s\tau}
    +\frac{1}{\beta}\sum_{s=-\infty}^{s=\infty}\frac{e^{-\im\omega_s\tau}}{\im\omega_s}.
  \end{split}
\end{equation}
The last term on the right-handed side is just the Fourier series of $-\frac{1}{2}$, thus we evaluate $\mathcal{G}_0(\tau)$ numerically as
\begin{equation}
  \mathcal{G}_0(\tau)=\frac{1}{\beta}\sum_{s=s_{\mathrm{min}}}^{s_{\mathrm{max}}}
  \qty[\mathcal{G}_0(\im\omega_s)-\frac{1}{\im\omega_s}]e^{-\im\omega_s\tau}-\frac{1}{2}.
\end{equation}
In this article we set $s_{\mathrm{min}}=-1001$ and $s_{\mathrm{max}}=1000$.

We denote the retarded Green's function in the noninteracting limit as $G_0(t)$, which is defined as
\begin{equation}
  \begin{split}
    G_0(t)=&-\im\theta(t)\expval*{\aop(t)\adop+\adop\aop(t)}\\
    =&-\frac{\im}{Z}\theta(t)\Tr\{e^{-\beta\Hop}[\aop(t)\adop+\adop\aop(t)]\},
  \end{split}
\end{equation}
where $\theta(t)$ is the Heaviside step function and $\aop(t)=e^{\im\Hop t}\aop e^{-\im\Hop t}$.
Its analytical solution in the frequency domain is
\begin{equation}
  G_0^{-1}(\omega)=\omega+\im0+\mu-\Delta(\omega),
\end{equation}
where $\im0$ is a positive imaginary infinitesimal \cite{mahan2000-many,lifshitz1980-statistical} and
\begin{equation}
  \Delta(\omega)=\int \frac{J(\varepsilon)}{\omega+\im0-\varepsilon}d\varepsilon
\end{equation}
is the hybridization function.
 

The asymptotic behavior of $G_0(\omega)$ is $G_0(\omega)=\frac{1}{\omega+\im0}$ as $\omega\to\pm\infty$. Then similarly, we can evaluate it numerically by a truncation in frequency with an asymptotic correction as
\begin{equation}
  G_0(t)=\int_{\omega_{\mathrm{min}}}^{\omega_{\mathrm{max}}}\qty[G(\omega)-\frac{1}{\omega+\im0}]e^{-\im\omega t}\frac{d\omega}{2\pi}-\im.
\end{equation}
In numerical evaluation, we use a small positive imaginary number ($10^{-8}\im$) to replace $\im0$ and the truncation is set as $\omega_{\mathrm{max}}=-\omega_{\mathrm{min}}=200\Gamma$.

The Green's functions calculated in this manner still contains a very small truncation error. We note that these Green's functions can be computed to machine precision using an adaptive quadrature method~\cite{kaye2022-libdlr,kaye2023-fast}.

\section{Self-consistency DMFT loop on the real-time axis}\label{app:dmft}
The CTQMC solver works in the imaginary-time axis, therefore the corresponding DMFT loop manipulates the Matsubara Green's function $\mathcal{G}(\tau)$. It takes $\mathcal{G}_0(\tau)$ or $\mathcal{G}_0(\im\omega_s)$ as the input to compute final Matsubara Green's function $\mathcal{G}(\tau)$ or $\mathcal{G}(\im\omega_s)$.

Once $\mathcal{G}$ in the impurity problem is obtained, we can use the Dyson equation in frequency domain \cite{GullWerner2011}
\begin{equation}
\mathcal{G}^{-1}(\im\omega_s)=\mathcal{G}_0(\im\omega_s)-\Sigma(\im\omega_s),
\end{equation}
to get the self-energy $\Sigma(\im\omega_s)$. With this self-energy, the new lattice Green's function $\mathcal{G}_{\mathrm{l}}$ can be obtained via
\begin{equation}
  \mathcal{G}_{\mathrm{l}}(\im\omega_s)=\int \frac{A_0(\varepsilon)}{\im\omega_s+\mu-\varepsilon-\Sigma(\im\omega_s)}d\varepsilon,
\end{equation}
where $A_0(\varepsilon)$ is the unperturbed density of states in the lattice and $\mu$ is the chemical potential. The new hybridization function $\Delta(\im\omega_s)$ can be obtained via
\begin{equation}
  \mathcal{G}_{\mathrm{l}}^{-1}(i\omega_s)=\im\omega_s+\mu-\Sigma(\im\omega_s)-\Delta(\im\omega_s).
\end{equation}
With this new $\Delta(\im\omega_s)$ we can generate a new $\mathcal{G}_0(\im\omega_s)$ to start a new loop,
and the DMFT loop terminates when Matsubara Green's function $\mathcal{G}$ converges.

In our formalism, we can directly work in the real time and frequency domain, therefore we focus on the retarded Green's function $G(t)$ and $G(\omega)$. Given a bath spectrum density $J(\varepsilon)$, we can calculate $G(t)$ via GTEMPO method, and $G(\omega)$ can be obtained via Fourier transform. The self-energy can be extracted by the Dyson equation 
\begin{equation}
  G^{-1}(\omega)=G_0^{-1}(\omega)-\Sigma(\omega),
\end{equation}
With this self-energy, the new lattice retarded Green's function can be obtained via
\begin{equation}
  G_{\mathrm{l}}(\omega)=\int \frac{A_0(\varepsilon)}{\omega+\im0+\mu-\varepsilon-\Sigma(\omega)}d\varepsilon.
\end{equation}
The new hybridization function $\Delta(\omega)$ is then obtained via
\begin{equation}
  G_{\mathrm{l}}^{-1}(\omega)=\omega+\im0+\mu-\Sigma(\omega)-\Delta(\omega).
\end{equation}

We need the bath spectrum density $J(\omega)$, rather than the hybridization function $\Delta(\omega)$, to start a new loop by the GTEMPO solver. By definition, the hybridization function $\Delta(\omega)$ can be expressed in terms of $J(\omega)$ as
\begin{equation}
  \Delta(\omega)=\int \frac{J(\varepsilon)}{\omega+\im0-\varepsilon}d\varepsilon.
\end{equation}
Recall that 
\begin{equation}
  \frac{1}{x+\im0}=P\frac{1}{x}-\im\pi\delta(x),
\end{equation}
where $P$ denotes the Cauchy principal value. The hybridization function is then
\begin{equation}
  \Delta(\omega)=P\int \frac{J(\varepsilon)}{\omega-\varepsilon}d\varepsilon-
  \im\pi J(\omega),
\end{equation}
which means 
\begin{equation}
  J(\omega)=-\frac{1}{\pi}\Im \Delta(\omega).
\end{equation}
With this new $J(\omega)$ we can start a new DMFT loop, and the loop terminates when $G$ converges.

\bibliography{refs}

\begin{thebibliography}{73}%
\makeatletter
\providecommand \@ifxundefined [1]{%
 \@ifx{#1\undefined}
}%
\providecommand \@ifnum [1]{%
 \ifnum #1\expandafter \@firstoftwo
 \else \expandafter \@secondoftwo
 \fi
}%
\providecommand \@ifx [1]{%
 \ifx #1\expandafter \@firstoftwo
 \else \expandafter \@secondoftwo
 \fi
}%
\providecommand \natexlab [1]{#1}%
\providecommand \enquote  [1]{``#1''}%
\providecommand \bibnamefont  [1]{#1}%
\providecommand \bibfnamefont [1]{#1}%
\providecommand \citenamefont [1]{#1}%
\providecommand \href@noop [0]{\@secondoftwo}%
\providecommand \href [0]{\begingroup \@sanitize@url \@href}%
\providecommand \@href[1]{\@@startlink{#1}\@@href}%
\providecommand \@@href[1]{\endgroup#1\@@endlink}%
\providecommand \@sanitize@url [0]{\catcode `\\12\catcode `\$12\catcode
  `\&12\catcode `\#12\catcode `\^12\catcode `\_12\catcode `\%12\relax}%
\providecommand \@@startlink[1]{}%
\providecommand \@@endlink[0]{}%
\providecommand \url  [0]{\begingroup\@sanitize@url \@url }%
\providecommand \@url [1]{\endgroup\@href {#1}{\urlprefix }}%
\providecommand \urlprefix  [0]{URL }%
\providecommand \Eprint [0]{\href }%
\providecommand \doibase [0]{https://doi.org/}%
\providecommand \selectlanguage [0]{\@gobble}%
\providecommand \bibinfo  [0]{\@secondoftwo}%
\providecommand \bibfield  [0]{\@secondoftwo}%
\providecommand \translation [1]{[#1]}%
\providecommand \BibitemOpen [0]{}%
\providecommand \bibitemStop [0]{}%
\providecommand \bibitemNoStop [0]{.\EOS\space}%
\providecommand \EOS [0]{\spacefactor3000\relax}%
\providecommand \BibitemShut  [1]{\csname bibitem#1\endcsname}%
\let\auto@bib@innerbib\@empty
\bibitem [{\citenamefont {Mahan}(2000)}]{mahan2000-many}%
  \BibitemOpen
  \bibfield  {author} {\bibinfo {author} {\bibfnamefont {G.~D.}\ \bibnamefont
  {Mahan}},\ }\href@noop {} {\emph {\bibinfo {title} {Many-Particle Physics}}}\
  (\bibinfo  {publisher} {Springer; 3nd edition},\ \bibinfo {year}
  {2000})\BibitemShut {NoStop}%
\bibitem [{\citenamefont {Weiss}(1993)}]{weiss1993-quantum}%
  \BibitemOpen
  \bibfield  {author} {\bibinfo {author} {\bibfnamefont {U.}~\bibnamefont
  {Weiss}},\ }\href {https://doi.org/10.1142/8334} {\emph {\bibinfo {title}
  {Quantum Dissipative Systems}}}\ (\bibinfo  {publisher} {World Scientific},\
  \bibinfo {address} {Singapore},\ \bibinfo {year} {1993})\BibitemShut
  {NoStop}%
\bibitem [{\citenamefont {Georges}\ \emph {et~al.}(1996)\citenamefont
  {Georges}, \citenamefont {Kotliar}, \citenamefont {Krauth},\ and\
  \citenamefont {Rozenberg}}]{georges1996-dynamical}%
  \BibitemOpen
  \bibfield  {author} {\bibinfo {author} {\bibfnamefont {A.}~\bibnamefont
  {Georges}}, \bibinfo {author} {\bibfnamefont {G.}~\bibnamefont {Kotliar}},
  \bibinfo {author} {\bibfnamefont {W.}~\bibnamefont {Krauth}},\ and\ \bibinfo
  {author} {\bibfnamefont {M.~J.}\ \bibnamefont {Rozenberg}},\ }\bibfield
  {title} {\bibinfo {title} {Dynamical mean-field theory of strongly correlated
  fermion systems and the limit of infinite dimensions},\ }\href
  {https://doi.org/10.1103/revmodphys.68.13} {\bibfield  {journal} {\bibinfo
  {journal} {Rev. Mod. Phys.}\ }\textbf {\bibinfo {volume} {68}},\ \bibinfo
  {pages} {13} (\bibinfo {year} {1996})}\BibitemShut {NoStop}%
\bibitem [{\citenamefont {Gull}\ \emph {et~al.}(2011)\citenamefont {Gull},
  \citenamefont {Millis}, \citenamefont {Lichtenstein}, \citenamefont
  {Rubtsov}, \citenamefont {Troyer},\ and\ \citenamefont
  {Werner}}]{GullWerner2011}%
  \BibitemOpen
  \bibfield  {author} {\bibinfo {author} {\bibfnamefont {E.}~\bibnamefont
  {Gull}}, \bibinfo {author} {\bibfnamefont {A.~J.}\ \bibnamefont {Millis}},
  \bibinfo {author} {\bibfnamefont {A.~I.}\ \bibnamefont {Lichtenstein}},
  \bibinfo {author} {\bibfnamefont {A.~N.}\ \bibnamefont {Rubtsov}}, \bibinfo
  {author} {\bibfnamefont {M.}~\bibnamefont {Troyer}},\ and\ \bibinfo {author}
  {\bibfnamefont {P.}~\bibnamefont {Werner}},\ }\bibfield  {title} {\bibinfo
  {title} {Continuous-time monte carlo methods for quantum impurity models},\
  }\href {https://doi.org/10.1103/RevModPhys.83.349} {\bibfield  {journal}
  {\bibinfo  {journal} {Rev. Mod. Phys.}\ }\textbf {\bibinfo {volume} {83}},\
  \bibinfo {pages} {349} (\bibinfo {year} {2011})}\BibitemShut {NoStop}%
\bibitem [{\citenamefont {Caffarel}\ and\ \citenamefont
  {Krauth}(1994)}]{CaffarelKrauth1994}%
  \BibitemOpen
  \bibfield  {author} {\bibinfo {author} {\bibfnamefont {M.}~\bibnamefont
  {Caffarel}}\ and\ \bibinfo {author} {\bibfnamefont {W.}~\bibnamefont
  {Krauth}},\ }\bibfield  {title} {\bibinfo {title} {Exact diagonalization
  approach to correlated fermions in infinite dimensions: Mott transition and
  superconductivity},\ }\href {https://doi.org/10.1103/PhysRevLett.72.1545}
  {\bibfield  {journal} {\bibinfo  {journal} {Phys. Rev. Lett.}\ }\textbf
  {\bibinfo {volume} {72}},\ \bibinfo {pages} {1545} (\bibinfo {year}
  {1994})}\BibitemShut {NoStop}%
\bibitem [{\citenamefont {Koch}\ \emph {et~al.}(2008)\citenamefont {Koch},
  \citenamefont {Sangiovanni},\ and\ \citenamefont
  {Gunnarsson}}]{KochGunnarsson2008}%
  \BibitemOpen
  \bibfield  {author} {\bibinfo {author} {\bibfnamefont {E.}~\bibnamefont
  {Koch}}, \bibinfo {author} {\bibfnamefont {G.}~\bibnamefont {Sangiovanni}},\
  and\ \bibinfo {author} {\bibfnamefont {O.}~\bibnamefont {Gunnarsson}},\
  }\bibfield  {title} {\bibinfo {title} {Sum rules and bath parametrization for
  quantum cluster theories},\ }\href
  {https://doi.org/10.1103/PhysRevB.78.115102} {\bibfield  {journal} {\bibinfo
  {journal} {Phys. Rev. B}\ }\textbf {\bibinfo {volume} {78}},\ \bibinfo
  {pages} {115102} (\bibinfo {year} {2008})}\BibitemShut {NoStop}%
\bibitem [{\citenamefont {Granath}\ and\ \citenamefont
  {Strand}(2012)}]{GranathStrand2012}%
  \BibitemOpen
  \bibfield  {author} {\bibinfo {author} {\bibfnamefont {M.}~\bibnamefont
  {Granath}}\ and\ \bibinfo {author} {\bibfnamefont {H.~U.~R.}\ \bibnamefont
  {Strand}},\ }\bibfield  {title} {\bibinfo {title} {Distributional exact
  diagonalization formalism for quantum impurity models},\ }\href
  {https://doi.org/10.1103/PhysRevB.86.115111} {\bibfield  {journal} {\bibinfo
  {journal} {Phys. Rev. B}\ }\textbf {\bibinfo {volume} {86}},\ \bibinfo
  {pages} {115111} (\bibinfo {year} {2012})}\BibitemShut {NoStop}%
\bibitem [{\citenamefont {Lu}\ \emph {et~al.}(2014)\citenamefont {Lu},
  \citenamefont {H\"oppner}, \citenamefont {Gunnarsson},\ and\ \citenamefont
  {Haverkort}}]{LuHaverkort2014}%
  \BibitemOpen
  \bibfield  {author} {\bibinfo {author} {\bibfnamefont {Y.}~\bibnamefont
  {Lu}}, \bibinfo {author} {\bibfnamefont {M.}~\bibnamefont {H\"oppner}},
  \bibinfo {author} {\bibfnamefont {O.}~\bibnamefont {Gunnarsson}},\ and\
  \bibinfo {author} {\bibfnamefont {M.~W.}\ \bibnamefont {Haverkort}},\
  }\bibfield  {title} {\bibinfo {title} {Efficient real-frequency solver for
  dynamical mean-field theory},\ }\href
  {https://doi.org/10.1103/PhysRevB.90.085102} {\bibfield  {journal} {\bibinfo
  {journal} {Phys. Rev. B}\ }\textbf {\bibinfo {volume} {90}},\ \bibinfo
  {pages} {085102} (\bibinfo {year} {2014})}\BibitemShut {NoStop}%
\bibitem [{\citenamefont {Mejuto-Zaera}\ \emph {et~al.}(2020)\citenamefont
  {Mejuto-Zaera}, \citenamefont {Zepeda-N\'u\~nez}, \citenamefont {Lindsey},
  \citenamefont {Tubman}, \citenamefont {Whaley},\ and\ \citenamefont
  {Lin}}]{ZaeraLin2020}%
  \BibitemOpen
  \bibfield  {author} {\bibinfo {author} {\bibfnamefont {C.}~\bibnamefont
  {Mejuto-Zaera}}, \bibinfo {author} {\bibfnamefont {L.}~\bibnamefont
  {Zepeda-N\'u\~nez}}, \bibinfo {author} {\bibfnamefont {M.}~\bibnamefont
  {Lindsey}}, \bibinfo {author} {\bibfnamefont {N.}~\bibnamefont {Tubman}},
  \bibinfo {author} {\bibfnamefont {B.}~\bibnamefont {Whaley}},\ and\ \bibinfo
  {author} {\bibfnamefont {L.}~\bibnamefont {Lin}},\ }\bibfield  {title}
  {\bibinfo {title} {Efficient hybridization fitting for dynamical mean-field
  theory via semi-definite relaxation},\ }\href
  {https://doi.org/10.1103/PhysRevB.101.035143} {\bibfield  {journal} {\bibinfo
   {journal} {Phys. Rev. B}\ }\textbf {\bibinfo {volume} {101}},\ \bibinfo
  {pages} {035143} (\bibinfo {year} {2020})}\BibitemShut {NoStop}%
\bibitem [{\citenamefont {He}\ and\ \citenamefont {Lu}(2014)}]{HeLu2014}%
  \BibitemOpen
  \bibfield  {author} {\bibinfo {author} {\bibfnamefont {R.-Q.}\ \bibnamefont
  {He}}\ and\ \bibinfo {author} {\bibfnamefont {Z.-Y.}\ \bibnamefont {Lu}},\
  }\bibfield  {title} {\bibinfo {title} {Quantum renormalization groups based
  on natural orbitals},\ }\href {https://doi.org/10.1103/PhysRevB.89.085108}
  {\bibfield  {journal} {\bibinfo  {journal} {Phys. Rev. B}\ }\textbf {\bibinfo
  {volume} {89}},\ \bibinfo {pages} {085108} (\bibinfo {year}
  {2014})}\BibitemShut {NoStop}%
\bibitem [{\citenamefont {He}\ \emph {et~al.}(2015)\citenamefont {He},
  \citenamefont {Dai},\ and\ \citenamefont {Lu}}]{HeLu2015}%
  \BibitemOpen
  \bibfield  {author} {\bibinfo {author} {\bibfnamefont {R.-Q.}\ \bibnamefont
  {He}}, \bibinfo {author} {\bibfnamefont {J.}~\bibnamefont {Dai}},\ and\
  \bibinfo {author} {\bibfnamefont {Z.-Y.}\ \bibnamefont {Lu}},\ }\bibfield
  {title} {\bibinfo {title} {Natural orbitals renormalization group approach to
  the two-impurity kondo critical point},\ }\href
  {https://doi.org/10.1103/PhysRevB.91.155140} {\bibfield  {journal} {\bibinfo
  {journal} {Phys. Rev. B}\ }\textbf {\bibinfo {volume} {91}},\ \bibinfo
  {pages} {155140} (\bibinfo {year} {2015})}\BibitemShut {NoStop}%
\bibitem [{\citenamefont {Wilson}(1975)}]{Wilson1975}%
  \BibitemOpen
  \bibfield  {author} {\bibinfo {author} {\bibfnamefont {K.~G.}\ \bibnamefont
  {Wilson}},\ }\bibfield  {title} {\bibinfo {title} {The renormalization group:
  Critical phenomena and the kondo problem},\ }\href
  {https://doi.org/10.1103/RevModPhys.47.773} {\bibfield  {journal} {\bibinfo
  {journal} {Rev. Mod. Phys.}\ }\textbf {\bibinfo {volume} {47}},\ \bibinfo
  {pages} {773} (\bibinfo {year} {1975})}\BibitemShut {NoStop}%
\bibitem [{\citenamefont {Bulla}(1999)}]{Bulla1999}%
  \BibitemOpen
  \bibfield  {author} {\bibinfo {author} {\bibfnamefont {R.}~\bibnamefont
  {Bulla}},\ }\bibfield  {title} {\bibinfo {title} {Zero temperature
  metal-insulator transition in the infinite-dimensional hubbard model},\
  }\href {https://doi.org/10.1103/PhysRevLett.83.136} {\bibfield  {journal}
  {\bibinfo  {journal} {Phys. Rev. Lett.}\ }\textbf {\bibinfo {volume} {83}},\
  \bibinfo {pages} {136} (\bibinfo {year} {1999})}\BibitemShut {NoStop}%
\bibitem [{\citenamefont {Bulla}\ \emph {et~al.}(2008)\citenamefont {Bulla},
  \citenamefont {Costi},\ and\ \citenamefont {Pruschke}}]{BullaPruschke2008}%
  \BibitemOpen
  \bibfield  {author} {\bibinfo {author} {\bibfnamefont {R.}~\bibnamefont
  {Bulla}}, \bibinfo {author} {\bibfnamefont {T.~A.}\ \bibnamefont {Costi}},\
  and\ \bibinfo {author} {\bibfnamefont {T.}~\bibnamefont {Pruschke}},\
  }\bibfield  {title} {\bibinfo {title} {Numerical renormalization group method
  for quantum impurity systems},\ }\href
  {https://doi.org/10.1103/RevModPhys.80.395} {\bibfield  {journal} {\bibinfo
  {journal} {Rev. Mod. Phys.}\ }\textbf {\bibinfo {volume} {80}},\ \bibinfo
  {pages} {395} (\bibinfo {year} {2008})}\BibitemShut {NoStop}%
\bibitem [{\citenamefont {Anders}(2008)}]{Frithjof2008}%
  \BibitemOpen
  \bibfield  {author} {\bibinfo {author} {\bibfnamefont {F.~B.}\ \bibnamefont
  {Anders}},\ }\bibfield  {title} {\bibinfo {title} {A numerical
  renormalization group approach to non-equilibrium green functions for quantum
  impurity models},\ }\href {https://doi.org/10.1088/0953-8984/20/19/195216}
  {\bibfield  {journal} {\bibinfo  {journal} {J. Phys. Condens. Matter}\
  }\textbf {\bibinfo {volume} {20}},\ \bibinfo {pages} {195216} (\bibinfo
  {year} {2008})}\BibitemShut {NoStop}%
\bibitem [{\citenamefont {\ifmmode~\check{Z}\else \v{Z}\fi{}itko}\ and\
  \citenamefont {Pruschke}(2009)}]{ZitkoPruschke2009}%
  \BibitemOpen
  \bibfield  {author} {\bibinfo {author} {\bibfnamefont {R.}~\bibnamefont
  {\ifmmode~\check{Z}\else \v{Z}\fi{}itko}}\ and\ \bibinfo {author}
  {\bibfnamefont {T.}~\bibnamefont {Pruschke}},\ }\bibfield  {title} {\bibinfo
  {title} {Energy resolution and discretization artifacts in the numerical
  renormalization group},\ }\href {https://doi.org/10.1103/PhysRevB.79.085106}
  {\bibfield  {journal} {\bibinfo  {journal} {Phys. Rev. B}\ }\textbf {\bibinfo
  {volume} {79}},\ \bibinfo {pages} {085106} (\bibinfo {year}
  {2009})}\BibitemShut {NoStop}%
\bibitem [{\citenamefont {Deng}\ \emph {et~al.}(2013)\citenamefont {Deng},
  \citenamefont {Mravlje}, \citenamefont {\ifmmode~\check{Z}\else
  \v{Z}\fi{}itko}, \citenamefont {Ferrero}, \citenamefont {Kotliar},\ and\
  \citenamefont {Georges}}]{DengGeorges2013}%
  \BibitemOpen
  \bibfield  {author} {\bibinfo {author} {\bibfnamefont {X.}~\bibnamefont
  {Deng}}, \bibinfo {author} {\bibfnamefont {J.}~\bibnamefont {Mravlje}},
  \bibinfo {author} {\bibfnamefont {R.}~\bibnamefont {\ifmmode~\check{Z}\else
  \v{Z}\fi{}itko}}, \bibinfo {author} {\bibfnamefont {M.}~\bibnamefont
  {Ferrero}}, \bibinfo {author} {\bibfnamefont {G.}~\bibnamefont {Kotliar}},\
  and\ \bibinfo {author} {\bibfnamefont {A.}~\bibnamefont {Georges}},\
  }\bibfield  {title} {\bibinfo {title} {How bad metals turn good:
  Spectroscopic signatures of resilient quasiparticles},\ }\href
  {https://doi.org/10.1103/PhysRevLett.110.086401} {\bibfield  {journal}
  {\bibinfo  {journal} {Phys. Rev. Lett.}\ }\textbf {\bibinfo {volume} {110}},\
  \bibinfo {pages} {086401} (\bibinfo {year} {2013})}\BibitemShut {NoStop}%
\bibitem [{\citenamefont {Stadler}\ \emph {et~al.}(2015)\citenamefont
  {Stadler}, \citenamefont {Yin}, \citenamefont {von Delft}, \citenamefont
  {Kotliar},\ and\ \citenamefont {Weichselbaum}}]{StadlerWeichselbaum2015}%
  \BibitemOpen
  \bibfield  {author} {\bibinfo {author} {\bibfnamefont {K.~M.}\ \bibnamefont
  {Stadler}}, \bibinfo {author} {\bibfnamefont {Z.~P.}\ \bibnamefont {Yin}},
  \bibinfo {author} {\bibfnamefont {J.}~\bibnamefont {von Delft}}, \bibinfo
  {author} {\bibfnamefont {G.}~\bibnamefont {Kotliar}},\ and\ \bibinfo {author}
  {\bibfnamefont {A.}~\bibnamefont {Weichselbaum}},\ }\bibfield  {title}
  {\bibinfo {title} {Dynamical mean-field theory plus numerical
  renormalization-group study of spin-orbital separation in a three-band hund
  metal},\ }\href {https://doi.org/10.1103/PhysRevLett.115.136401} {\bibfield
  {journal} {\bibinfo  {journal} {Phys. Rev. Lett.}\ }\textbf {\bibinfo
  {volume} {115}},\ \bibinfo {pages} {136401} (\bibinfo {year}
  {2015})}\BibitemShut {NoStop}%
\bibitem [{\citenamefont {Lee}\ and\ \citenamefont
  {Weichselbaum}(2016)}]{LeeWeichselbaum2016}%
  \BibitemOpen
  \bibfield  {author} {\bibinfo {author} {\bibfnamefont {S.-S.~B.}\
  \bibnamefont {Lee}}\ and\ \bibinfo {author} {\bibfnamefont {A.}~\bibnamefont
  {Weichselbaum}},\ }\bibfield  {title} {\bibinfo {title} {Adaptive broadening
  to improve spectral resolution in the numerical renormalization group},\
  }\href {https://doi.org/10.1103/PhysRevB.94.235127} {\bibfield  {journal}
  {\bibinfo  {journal} {Phys. Rev. B}\ }\textbf {\bibinfo {volume} {94}},\
  \bibinfo {pages} {235127} (\bibinfo {year} {2016})}\BibitemShut {NoStop}%
\bibitem [{\citenamefont {Lee}\ \emph {et~al.}(2017)\citenamefont {Lee},
  \citenamefont {von Delft},\ and\ \citenamefont
  {Weichselbaum}}]{LeeWeichselbaum2017}%
  \BibitemOpen
  \bibfield  {author} {\bibinfo {author} {\bibfnamefont {S.-S.~B.}\
  \bibnamefont {Lee}}, \bibinfo {author} {\bibfnamefont {J.}~\bibnamefont {von
  Delft}},\ and\ \bibinfo {author} {\bibfnamefont {A.}~\bibnamefont
  {Weichselbaum}},\ }\bibfield  {title} {\bibinfo {title} {Doublon-holon origin
  of the subpeaks at the hubbard band edges},\ }\href
  {https://doi.org/10.1103/PhysRevLett.119.236402} {\bibfield  {journal}
  {\bibinfo  {journal} {Phys. Rev. Lett.}\ }\textbf {\bibinfo {volume} {119}},\
  \bibinfo {pages} {236402} (\bibinfo {year} {2017})}\BibitemShut {NoStop}%
\bibitem [{\citenamefont {Wolf}\ \emph {et~al.}(2014)\citenamefont {Wolf},
  \citenamefont {McCulloch}, \citenamefont {Parcollet},\ and\ \citenamefont
  {Schollw\"ock}}]{WolfSchollwock2014b}%
  \BibitemOpen
  \bibfield  {author} {\bibinfo {author} {\bibfnamefont {F.~A.}\ \bibnamefont
  {Wolf}}, \bibinfo {author} {\bibfnamefont {I.~P.}\ \bibnamefont {McCulloch}},
  \bibinfo {author} {\bibfnamefont {O.}~\bibnamefont {Parcollet}},\ and\
  \bibinfo {author} {\bibfnamefont {U.}~\bibnamefont {Schollw\"ock}},\
  }\bibfield  {title} {\bibinfo {title} {Chebyshev matrix product state
  impurity solver for dynamical mean-field theory},\ }\href
  {https://doi.org/10.1103/PhysRevB.90.115124} {\bibfield  {journal} {\bibinfo
  {journal} {Phys. Rev. B}\ }\textbf {\bibinfo {volume} {90}},\ \bibinfo
  {pages} {115124} (\bibinfo {year} {2014})}\BibitemShut {NoStop}%
\bibitem [{\citenamefont {Ganahl}\ \emph {et~al.}(2014)\citenamefont {Ganahl},
  \citenamefont {Thunstr\"om}, \citenamefont {Verstraete}, \citenamefont
  {Held},\ and\ \citenamefont {Evertz}}]{GanahlEvertz2014}%
  \BibitemOpen
  \bibfield  {author} {\bibinfo {author} {\bibfnamefont {M.}~\bibnamefont
  {Ganahl}}, \bibinfo {author} {\bibfnamefont {P.}~\bibnamefont {Thunstr\"om}},
  \bibinfo {author} {\bibfnamefont {F.}~\bibnamefont {Verstraete}}, \bibinfo
  {author} {\bibfnamefont {K.}~\bibnamefont {Held}},\ and\ \bibinfo {author}
  {\bibfnamefont {H.~G.}\ \bibnamefont {Evertz}},\ }\bibfield  {title}
  {\bibinfo {title} {Chebyshev expansion for impurity models using matrix
  product states},\ }\href {https://doi.org/10.1103/PhysRevB.90.045144}
  {\bibfield  {journal} {\bibinfo  {journal} {Phys. Rev. B}\ }\textbf {\bibinfo
  {volume} {90}},\ \bibinfo {pages} {045144} (\bibinfo {year}
  {2014})}\BibitemShut {NoStop}%
\bibitem [{\citenamefont {Ganahl}\ \emph {et~al.}(2015)\citenamefont {Ganahl},
  \citenamefont {Aichhorn}, \citenamefont {Evertz}, \citenamefont
  {Thunstr\"om}, \citenamefont {Held},\ and\ \citenamefont
  {Verstraete}}]{GanahlVerstraete2015}%
  \BibitemOpen
  \bibfield  {author} {\bibinfo {author} {\bibfnamefont {M.}~\bibnamefont
  {Ganahl}}, \bibinfo {author} {\bibfnamefont {M.}~\bibnamefont {Aichhorn}},
  \bibinfo {author} {\bibfnamefont {H.~G.}\ \bibnamefont {Evertz}}, \bibinfo
  {author} {\bibfnamefont {P.}~\bibnamefont {Thunstr\"om}}, \bibinfo {author}
  {\bibfnamefont {K.}~\bibnamefont {Held}},\ and\ \bibinfo {author}
  {\bibfnamefont {F.}~\bibnamefont {Verstraete}},\ }\bibfield  {title}
  {\bibinfo {title} {Efficient dmft impurity solver using real-time dynamics
  with matrix product states},\ }\href
  {https://doi.org/10.1103/PhysRevB.92.155132} {\bibfield  {journal} {\bibinfo
  {journal} {Phys. Rev. B}\ }\textbf {\bibinfo {volume} {92}},\ \bibinfo
  {pages} {155132} (\bibinfo {year} {2015})}\BibitemShut {NoStop}%
\bibitem [{\citenamefont {Wolf}\ \emph {et~al.}(2015)\citenamefont {Wolf},
  \citenamefont {Go}, \citenamefont {McCulloch}, \citenamefont {Millis},\ and\
  \citenamefont {Schollw\"ock}}]{WolfSchollwock2015}%
  \BibitemOpen
  \bibfield  {author} {\bibinfo {author} {\bibfnamefont {F.~A.}\ \bibnamefont
  {Wolf}}, \bibinfo {author} {\bibfnamefont {A.}~\bibnamefont {Go}}, \bibinfo
  {author} {\bibfnamefont {I.~P.}\ \bibnamefont {McCulloch}}, \bibinfo {author}
  {\bibfnamefont {A.~J.}\ \bibnamefont {Millis}},\ and\ \bibinfo {author}
  {\bibfnamefont {U.}~\bibnamefont {Schollw\"ock}},\ }\bibfield  {title}
  {\bibinfo {title} {Imaginary-time matrix product state impurity solver for
  dynamical mean-field theory},\ }\href
  {https://doi.org/10.1103/PhysRevX.5.041032} {\bibfield  {journal} {\bibinfo
  {journal} {Phys. Rev. X}\ }\textbf {\bibinfo {volume} {5}},\ \bibinfo {pages}
  {041032} (\bibinfo {year} {2015})}\BibitemShut {NoStop}%
\bibitem [{\citenamefont {Garc\'{\i}a}\ \emph {et~al.}(2004)\citenamefont
  {Garc\'{\i}a}, \citenamefont {Hallberg},\ and\ \citenamefont
  {Rozenberg}}]{GarciaRozenberg2004}%
  \BibitemOpen
  \bibfield  {author} {\bibinfo {author} {\bibfnamefont {D.~J.}\ \bibnamefont
  {Garc\'{\i}a}}, \bibinfo {author} {\bibfnamefont {K.}~\bibnamefont
  {Hallberg}},\ and\ \bibinfo {author} {\bibfnamefont {M.~J.}\ \bibnamefont
  {Rozenberg}},\ }\bibfield  {title} {\bibinfo {title} {Dynamical mean field
  theory with the density matrix renormalization group},\ }\href
  {https://doi.org/10.1103/PhysRevLett.93.246403} {\bibfield  {journal}
  {\bibinfo  {journal} {Phys. Rev. Lett.}\ }\textbf {\bibinfo {volume} {93}},\
  \bibinfo {pages} {246403} (\bibinfo {year} {2004})}\BibitemShut {NoStop}%
\bibitem [{\citenamefont {Nishimoto}\ \emph {et~al.}(2006)\citenamefont
  {Nishimoto}, \citenamefont {Gebhard},\ and\ \citenamefont
  {Jeckelmann}}]{NishimotoJeckelmann2006}%
  \BibitemOpen
  \bibfield  {author} {\bibinfo {author} {\bibfnamefont {S.}~\bibnamefont
  {Nishimoto}}, \bibinfo {author} {\bibfnamefont {F.}~\bibnamefont {Gebhard}},\
  and\ \bibinfo {author} {\bibfnamefont {E.}~\bibnamefont {Jeckelmann}},\
  }\bibfield  {title} {\bibinfo {title} {Dynamical mean-field theory
  calculation with the dynamical density-matrix renormalization group},\ }\href
  {https://doi.org/https://doi.org/10.1016/j.physb.2006.01.104} {\bibfield
  {journal} {\bibinfo  {journal} {Physica B Condens. Matter}\ }\textbf
  {\bibinfo {volume} {378-380}},\ \bibinfo {pages} {283} (\bibinfo {year}
  {2006})}\BibitemShut {NoStop}%
\bibitem [{\citenamefont {Weichselbaum}\ \emph {et~al.}(2009)\citenamefont
  {Weichselbaum}, \citenamefont {Verstraete}, \citenamefont {Schollw\"ock},
  \citenamefont {Cirac},\ and\ \citenamefont {von
  Delft}}]{WeichselbaumDelft2009}%
  \BibitemOpen
  \bibfield  {author} {\bibinfo {author} {\bibfnamefont {A.}~\bibnamefont
  {Weichselbaum}}, \bibinfo {author} {\bibfnamefont {F.}~\bibnamefont
  {Verstraete}}, \bibinfo {author} {\bibfnamefont {U.}~\bibnamefont
  {Schollw\"ock}}, \bibinfo {author} {\bibfnamefont {J.~I.}\ \bibnamefont
  {Cirac}},\ and\ \bibinfo {author} {\bibfnamefont {J.}~\bibnamefont {von
  Delft}},\ }\bibfield  {title} {\bibinfo {title} {Variational
  matrix-product-state approach to quantum impurity models},\ }\href
  {https://doi.org/10.1103/PhysRevB.80.165117} {\bibfield  {journal} {\bibinfo
  {journal} {Phys. Rev. B}\ }\textbf {\bibinfo {volume} {80}},\ \bibinfo
  {pages} {165117} (\bibinfo {year} {2009})}\BibitemShut {NoStop}%
\bibitem [{\citenamefont {Bauernfeind}\ \emph {et~al.}(2017)\citenamefont
  {Bauernfeind}, \citenamefont {Zingl}, \citenamefont {Triebl}, \citenamefont
  {Aichhorn},\ and\ \citenamefont {Evertz}}]{BauernfeindEvertz2017}%
  \BibitemOpen
  \bibfield  {author} {\bibinfo {author} {\bibfnamefont {D.}~\bibnamefont
  {Bauernfeind}}, \bibinfo {author} {\bibfnamefont {M.}~\bibnamefont {Zingl}},
  \bibinfo {author} {\bibfnamefont {R.}~\bibnamefont {Triebl}}, \bibinfo
  {author} {\bibfnamefont {M.}~\bibnamefont {Aichhorn}},\ and\ \bibinfo
  {author} {\bibfnamefont {H.~G.}\ \bibnamefont {Evertz}},\ }\bibfield  {title}
  {\bibinfo {title} {Fork tensor-product states: Efficient multiorbital
  real-time dmft solver},\ }\href {https://doi.org/10.1103/PhysRevX.7.031013}
  {\bibfield  {journal} {\bibinfo  {journal} {Phys. Rev. X}\ }\textbf {\bibinfo
  {volume} {7}},\ \bibinfo {pages} {031013} (\bibinfo {year}
  {2017})}\BibitemShut {NoStop}%
\bibitem [{\citenamefont {Lu}\ \emph {et~al.}(2019)\citenamefont {Lu},
  \citenamefont {Cao}, \citenamefont {Hansmann},\ and\ \citenamefont
  {Haverkort}}]{LuHaverkort2019}%
  \BibitemOpen
  \bibfield  {author} {\bibinfo {author} {\bibfnamefont {Y.}~\bibnamefont
  {Lu}}, \bibinfo {author} {\bibfnamefont {X.}~\bibnamefont {Cao}}, \bibinfo
  {author} {\bibfnamefont {P.}~\bibnamefont {Hansmann}},\ and\ \bibinfo
  {author} {\bibfnamefont {M.~W.}\ \bibnamefont {Haverkort}},\ }\bibfield
  {title} {\bibinfo {title} {Natural-orbital impurity solver and projection
  approach for green's functions},\ }\href
  {https://doi.org/10.1103/PhysRevB.100.115134} {\bibfield  {journal} {\bibinfo
   {journal} {Phys. Rev. B}\ }\textbf {\bibinfo {volume} {100}},\ \bibinfo
  {pages} {115134} (\bibinfo {year} {2019})}\BibitemShut {NoStop}%
\bibitem [{\citenamefont {Werner}\ \emph {et~al.}(2023)\citenamefont {Werner},
  \citenamefont {Lotze},\ and\ \citenamefont {Arrigoni}}]{WernerArrigoni2023}%
  \BibitemOpen
  \bibfield  {author} {\bibinfo {author} {\bibfnamefont {D.}~\bibnamefont
  {Werner}}, \bibinfo {author} {\bibfnamefont {J.}~\bibnamefont {Lotze}},\ and\
  \bibinfo {author} {\bibfnamefont {E.}~\bibnamefont {Arrigoni}},\ }\bibfield
  {title} {\bibinfo {title} {Configuration interaction based nonequilibrium
  steady state impurity solver},\ }\href
  {https://doi.org/10.1103/PhysRevB.107.075119} {\bibfield  {journal} {\bibinfo
   {journal} {Phys. Rev. B}\ }\textbf {\bibinfo {volume} {107}},\ \bibinfo
  {pages} {075119} (\bibinfo {year} {2023})}\BibitemShut {NoStop}%
\bibitem [{\citenamefont {Kohn}\ and\ \citenamefont
  {Santoro}(2021)}]{KohnSantoro2021}%
  \BibitemOpen
  \bibfield  {author} {\bibinfo {author} {\bibfnamefont {L.}~\bibnamefont
  {Kohn}}\ and\ \bibinfo {author} {\bibfnamefont {G.~E.}\ \bibnamefont
  {Santoro}},\ }\bibfield  {title} {\bibinfo {title} {Efficient mapping for
  anderson impurity problems with matrix product states},\ }\href
  {https://doi.org/10.1103/PhysRevB.104.014303} {\bibfield  {journal} {\bibinfo
   {journal} {Phys. Rev. B}\ }\textbf {\bibinfo {volume} {104}},\ \bibinfo
  {pages} {014303} (\bibinfo {year} {2021})}\BibitemShut {NoStop}%
\bibitem [{\citenamefont {Kohn}\ and\ \citenamefont
  {Santoro}(2022)}]{KohnSantoro2022}%
  \BibitemOpen
  \bibfield  {author} {\bibinfo {author} {\bibfnamefont {L.}~\bibnamefont
  {Kohn}}\ and\ \bibinfo {author} {\bibfnamefont {G.~E.}\ \bibnamefont
  {Santoro}},\ }\bibfield  {title} {\bibinfo {title} {Quench dynamics of the
  anderson impurity model at finite temperature using matrix product states:
  entanglement and bath dynamics},\ }\href
  {https://doi.org/10.1088/1742-5468/ac729b} {\bibfield  {journal} {\bibinfo
  {journal} {J. Stat. Mech. Theory Exp.}\ }\textbf {\bibinfo {volume} {2022}},\
  \bibinfo {pages} {063102} (\bibinfo {year} {2022})}\BibitemShut {NoStop}%
\bibitem [{\citenamefont {Feynman}\ and\ \citenamefont
  {Vernon}(1963)}]{FeynmanVernon1963}%
  \BibitemOpen
  \bibfield  {author} {\bibinfo {author} {\bibfnamefont {R.~P.}\ \bibnamefont
  {Feynman}}\ and\ \bibinfo {author} {\bibfnamefont {F.~L.}\ \bibnamefont
  {Vernon}},\ }\bibfield  {title} {\bibinfo {title} {The theory of a general
  quantum system interacting with a linear dissipative system},\ }\href
  {https://doi.org/10.1016/0003-4916(63)90068-X} {\bibfield  {journal}
  {\bibinfo  {journal} {Ann. Phys.}\ }\textbf {\bibinfo {volume} {24}},\
  \bibinfo {pages} {118} (\bibinfo {year} {1963})}\BibitemShut {NoStop}%
\bibitem [{\citenamefont {Rubtsov}\ \emph {et~al.}(2005)\citenamefont
  {Rubtsov}, \citenamefont {Savkin},\ and\ \citenamefont
  {Lichtenstein}}]{RubtsovLichtenstein2005}%
  \BibitemOpen
  \bibfield  {author} {\bibinfo {author} {\bibfnamefont {A.~N.}\ \bibnamefont
  {Rubtsov}}, \bibinfo {author} {\bibfnamefont {V.~V.}\ \bibnamefont
  {Savkin}},\ and\ \bibinfo {author} {\bibfnamefont {A.~I.}\ \bibnamefont
  {Lichtenstein}},\ }\bibfield  {title} {\bibinfo {title} {Continuous-time
  quantum monte carlo method for fermions},\ }\href
  {https://doi.org/10.1103/PhysRevB.72.035122} {\bibfield  {journal} {\bibinfo
  {journal} {Phys. Rev. B}\ }\textbf {\bibinfo {volume} {72}},\ \bibinfo
  {pages} {035122} (\bibinfo {year} {2005})}\BibitemShut {NoStop}%
\bibitem [{\citenamefont {Gull}\ \emph {et~al.}(2008)\citenamefont {Gull},
  \citenamefont {Werner}, \citenamefont {Parcollet},\ and\ \citenamefont
  {Troyer}}]{GullTroyer2008}%
  \BibitemOpen
  \bibfield  {author} {\bibinfo {author} {\bibfnamefont {E.}~\bibnamefont
  {Gull}}, \bibinfo {author} {\bibfnamefont {P.}~\bibnamefont {Werner}},
  \bibinfo {author} {\bibfnamefont {O.}~\bibnamefont {Parcollet}},\ and\
  \bibinfo {author} {\bibfnamefont {M.}~\bibnamefont {Troyer}},\ }\bibfield
  {title} {\bibinfo {title} {Continuous-time auxiliary-field monte carlo for
  quantum impurity models},\ }\href
  {https://doi.org/10.1209/0295-5075/82/57003} {\bibfield  {journal} {\bibinfo
  {journal} {EPL}\ }\textbf {\bibinfo {volume} {82}},\ \bibinfo {pages} {57003}
  (\bibinfo {year} {2008})}\BibitemShut {NoStop}%
\bibitem [{\citenamefont {Werner}\ and\ \citenamefont
  {Millis}(2006)}]{WernerMillis2006b}%
  \BibitemOpen
  \bibfield  {author} {\bibinfo {author} {\bibfnamefont {P.}~\bibnamefont
  {Werner}}\ and\ \bibinfo {author} {\bibfnamefont {A.~J.}\ \bibnamefont
  {Millis}},\ }\bibfield  {title} {\bibinfo {title} {Hybridization expansion
  impurity solver: General formulation and application to kondo lattice and
  two-orbital models},\ }\href {https://doi.org/10.1103/PhysRevB.74.155107}
  {\bibfield  {journal} {\bibinfo  {journal} {Phys. Rev. B}\ }\textbf {\bibinfo
  {volume} {74}},\ \bibinfo {pages} {155107} (\bibinfo {year}
  {2006})}\BibitemShut {NoStop}%
\bibitem [{\citenamefont {Werner}\ \emph {et~al.}(2006)\citenamefont {Werner},
  \citenamefont {Comanac}, \citenamefont {de' Medici}, \citenamefont {Troyer},\
  and\ \citenamefont {Millis}}]{WernerMillis2006}%
  \BibitemOpen
  \bibfield  {author} {\bibinfo {author} {\bibfnamefont {P.}~\bibnamefont
  {Werner}}, \bibinfo {author} {\bibfnamefont {A.}~\bibnamefont {Comanac}},
  \bibinfo {author} {\bibfnamefont {L.}~\bibnamefont {de' Medici}}, \bibinfo
  {author} {\bibfnamefont {M.}~\bibnamefont {Troyer}},\ and\ \bibinfo {author}
  {\bibfnamefont {A.~J.}\ \bibnamefont {Millis}},\ }\bibfield  {title}
  {\bibinfo {title} {Continuous-time solver for quantum impurity models},\
  }\href {https://doi.org/10.1103/PhysRevLett.97.076405} {\bibfield  {journal}
  {\bibinfo  {journal} {Phys. Rev. Lett.}\ }\textbf {\bibinfo {volume} {97}},\
  \bibinfo {pages} {076405} (\bibinfo {year} {2006})}\BibitemShut {NoStop}%
\bibitem [{\citenamefont {Shinaoka}\ \emph {et~al.}(2017)\citenamefont
  {Shinaoka}, \citenamefont {Gull},\ and\ \citenamefont
  {Werner}}]{ShinaokaWerner2017}%
  \BibitemOpen
  \bibfield  {author} {\bibinfo {author} {\bibfnamefont {H.}~\bibnamefont
  {Shinaoka}}, \bibinfo {author} {\bibfnamefont {E.}~\bibnamefont {Gull}},\
  and\ \bibinfo {author} {\bibfnamefont {P.}~\bibnamefont {Werner}},\
  }\bibfield  {title} {\bibinfo {title} {Continuous-time hybridization
  expansion quantum impurity solver for multi-orbital systems with complex
  hybridizations},\ }\href
  {https://doi.org/https://doi.org/10.1016/j.cpc.2017.01.003} {\bibfield
  {journal} {\bibinfo  {journal} {Comput. Phys. Commun.}\ }\textbf {\bibinfo
  {volume} {215}},\ \bibinfo {pages} {128} (\bibinfo {year}
  {2017})}\BibitemShut {NoStop}%
\bibitem [{\citenamefont {Eidelstein}\ \emph {et~al.}(2020)\citenamefont
  {Eidelstein}, \citenamefont {Gull},\ and\ \citenamefont
  {Cohen}}]{EidelsteinCohen2020}%
  \BibitemOpen
  \bibfield  {author} {\bibinfo {author} {\bibfnamefont {E.}~\bibnamefont
  {Eidelstein}}, \bibinfo {author} {\bibfnamefont {E.}~\bibnamefont {Gull}},\
  and\ \bibinfo {author} {\bibfnamefont {G.}~\bibnamefont {Cohen}},\ }\bibfield
   {title} {\bibinfo {title} {Multiorbital quantum impurity solver for general
  interactions and hybridizations},\ }\href
  {https://doi.org/10.1103/PhysRevLett.124.206405} {\bibfield  {journal}
  {\bibinfo  {journal} {Phys. Rev. Lett.}\ }\textbf {\bibinfo {volume} {124}},\
  \bibinfo {pages} {206405} (\bibinfo {year} {2020})}\BibitemShut {NoStop}%
\bibitem [{\citenamefont {Cohen}\ \emph
  {et~al.}(2014{\natexlab{a}})\citenamefont {Cohen}, \citenamefont {Reichman},
  \citenamefont {Millis},\ and\ \citenamefont {Gull}}]{CohenGull2014}%
  \BibitemOpen
  \bibfield  {author} {\bibinfo {author} {\bibfnamefont {G.}~\bibnamefont
  {Cohen}}, \bibinfo {author} {\bibfnamefont {D.~R.}\ \bibnamefont {Reichman}},
  \bibinfo {author} {\bibfnamefont {A.~J.}\ \bibnamefont {Millis}},\ and\
  \bibinfo {author} {\bibfnamefont {E.}~\bibnamefont {Gull}},\ }\bibfield
  {title} {\bibinfo {title} {Green's functions from real-time bold-line monte
  carlo},\ }\href {https://doi.org/10.1103/PhysRevB.89.115139} {\bibfield
  {journal} {\bibinfo  {journal} {Phys. Rev. B}\ }\textbf {\bibinfo {volume}
  {89}},\ \bibinfo {pages} {115139} (\bibinfo {year}
  {2014}{\natexlab{a}})}\BibitemShut {NoStop}%
\bibitem [{\citenamefont {Cohen}\ \emph
  {et~al.}(2014{\natexlab{b}})\citenamefont {Cohen}, \citenamefont {Gull},
  \citenamefont {Reichman},\ and\ \citenamefont {Millis}}]{CohenMillis2014}%
  \BibitemOpen
  \bibfield  {author} {\bibinfo {author} {\bibfnamefont {G.}~\bibnamefont
  {Cohen}}, \bibinfo {author} {\bibfnamefont {E.}~\bibnamefont {Gull}},
  \bibinfo {author} {\bibfnamefont {D.~R.}\ \bibnamefont {Reichman}},\ and\
  \bibinfo {author} {\bibfnamefont {A.~J.}\ \bibnamefont {Millis}},\ }\bibfield
   {title} {\bibinfo {title} {Green's functions from real-time bold-line monte
  carlo calculations: Spectral properties of the nonequilibrium anderson
  impurity model},\ }\href {https://doi.org/10.1103/PhysRevLett.112.146802}
  {\bibfield  {journal} {\bibinfo  {journal} {Phys. Rev. Lett.}\ }\textbf
  {\bibinfo {volume} {112}},\ \bibinfo {pages} {146802} (\bibinfo {year}
  {2014}{\natexlab{b}})}\BibitemShut {NoStop}%
\bibitem [{\citenamefont {Cohen}\ \emph {et~al.}(2015)\citenamefont {Cohen},
  \citenamefont {Gull}, \citenamefont {Reichman},\ and\ \citenamefont
  {Millis}}]{CohenMillis2015}%
  \BibitemOpen
  \bibfield  {author} {\bibinfo {author} {\bibfnamefont {G.}~\bibnamefont
  {Cohen}}, \bibinfo {author} {\bibfnamefont {E.}~\bibnamefont {Gull}},
  \bibinfo {author} {\bibfnamefont {D.~R.}\ \bibnamefont {Reichman}},\ and\
  \bibinfo {author} {\bibfnamefont {A.~J.}\ \bibnamefont {Millis}},\ }\bibfield
   {title} {\bibinfo {title} {Taming the dynamical sign problem in real-time
  evolution of quantum many-body problems},\ }\href
  {https://doi.org/10.1103/PhysRevLett.115.266802} {\bibfield  {journal}
  {\bibinfo  {journal} {Phys. Rev. Lett.}\ }\textbf {\bibinfo {volume} {115}},\
  \bibinfo {pages} {266802} (\bibinfo {year} {2015})}\BibitemShut {NoStop}%
\bibitem [{\citenamefont {Chen}\ \emph
  {et~al.}(2017{\natexlab{a}})\citenamefont {Chen}, \citenamefont {Cohen},\
  and\ \citenamefont {Reichman}}]{ChenReichman2017a}%
  \BibitemOpen
  \bibfield  {author} {\bibinfo {author} {\bibfnamefont {H.-T.}\ \bibnamefont
  {Chen}}, \bibinfo {author} {\bibfnamefont {G.}~\bibnamefont {Cohen}},\ and\
  \bibinfo {author} {\bibfnamefont {D.~R.}\ \bibnamefont {Reichman}},\
  }\bibfield  {title} {\bibinfo {title} {{Inchworm Monte Carlo for exact
  non-adiabatic dynamics. I. Theory and algorithms}},\ }\href
  {https://doi.org/10.1063/1.4974328} {\bibfield  {journal} {\bibinfo
  {journal} {J. Chem. Phys.}\ }\textbf {\bibinfo {volume} {146}},\ \bibinfo
  {pages} {054105} (\bibinfo {year} {2017}{\natexlab{a}})}\BibitemShut
  {NoStop}%
\bibitem [{\citenamefont {Chen}\ \emph
  {et~al.}(2017{\natexlab{b}})\citenamefont {Chen}, \citenamefont {Cohen},\
  and\ \citenamefont {Reichman}}]{ChenReichman2017b}%
  \BibitemOpen
  \bibfield  {author} {\bibinfo {author} {\bibfnamefont {H.-T.}\ \bibnamefont
  {Chen}}, \bibinfo {author} {\bibfnamefont {G.}~\bibnamefont {Cohen}},\ and\
  \bibinfo {author} {\bibfnamefont {D.~R.}\ \bibnamefont {Reichman}},\
  }\bibfield  {title} {\bibinfo {title} {{Inchworm Monte Carlo for exact
  non-adiabatic dynamics. II. Benchmarks and comparison with established
  methods}},\ }\href {https://doi.org/10.1063/1.4974329} {\bibfield  {journal}
  {\bibinfo  {journal} {J. Chem. Phys.}\ }\textbf {\bibinfo {volume} {146}},\
  \bibinfo {pages} {054106} (\bibinfo {year} {2017}{\natexlab{b}})}\BibitemShut
  {NoStop}%
\bibitem [{\citenamefont {Erpenbeck}\ \emph {et~al.}(2023)\citenamefont
  {Erpenbeck}, \citenamefont {Gull},\ and\ \citenamefont
  {Cohen}}]{ErpenbeckCohen2023}%
  \BibitemOpen
  \bibfield  {author} {\bibinfo {author} {\bibfnamefont {A.}~\bibnamefont
  {Erpenbeck}}, \bibinfo {author} {\bibfnamefont {E.}~\bibnamefont {Gull}},\
  and\ \bibinfo {author} {\bibfnamefont {G.}~\bibnamefont {Cohen}},\ }\bibfield
   {title} {\bibinfo {title} {Quantum monte carlo method in the steady state},\
  }\href {https://doi.org/10.1103/PhysRevLett.130.186301} {\bibfield  {journal}
  {\bibinfo  {journal} {Phys. Rev. Lett.}\ }\textbf {\bibinfo {volume} {130}},\
  \bibinfo {pages} {186301} (\bibinfo {year} {2023})}\BibitemShut {NoStop}%
\bibitem [{\citenamefont {Tanimura}\ and\ \citenamefont
  {Kubo}(1989)}]{YoshitakaKubo1989}%
  \BibitemOpen
  \bibfield  {author} {\bibinfo {author} {\bibfnamefont {Y.}~\bibnamefont
  {Tanimura}}\ and\ \bibinfo {author} {\bibfnamefont {R.}~\bibnamefont
  {Kubo}},\ }\bibfield  {title} {\bibinfo {title} {Time evolution of a quantum
  system in contact with a nearly gaussian-markoffian noise bath},\ }\href
  {https://doi.org/10.1143/JPSJ.58.101} {\bibfield  {journal} {\bibinfo
  {journal} {J. Phys. Soc. Jpn.}\ }\textbf {\bibinfo {volume} {58}},\ \bibinfo
  {pages} {101} (\bibinfo {year} {1989})}\BibitemShut {NoStop}%
\bibitem [{\citenamefont {Jin}\ \emph {et~al.}(2007)\citenamefont {Jin},
  \citenamefont {Welack}, \citenamefont {Luo}, \citenamefont {Li},
  \citenamefont {Cui}, \citenamefont {Xu},\ and\ \citenamefont
  {Yan}}]{jin2007-dynamics}%
  \BibitemOpen
  \bibfield  {author} {\bibinfo {author} {\bibfnamefont {J.}~\bibnamefont
  {Jin}}, \bibinfo {author} {\bibfnamefont {S.}~\bibnamefont {Welack}},
  \bibinfo {author} {\bibfnamefont {J.}~\bibnamefont {Luo}}, \bibinfo {author}
  {\bibfnamefont {X.-Q.}\ \bibnamefont {Li}}, \bibinfo {author} {\bibfnamefont
  {P.}~\bibnamefont {Cui}}, \bibinfo {author} {\bibfnamefont {R.-X.}\
  \bibnamefont {Xu}},\ and\ \bibinfo {author} {\bibfnamefont {Y.}~\bibnamefont
  {Yan}},\ }\bibfield  {title} {\bibinfo {title} {Dynamics of quantum
  dissipation systems interacting with fermion and boson grand canonical bath
  ensembles: Hierarchical equations of motion approach},\ }\href
  {https://doi.org/10.1063/1.2713104} {\bibfield  {journal} {\bibinfo
  {journal} {J. Chem. Phys.}\ }\textbf {\bibinfo {volume} {126}},\ \bibinfo
  {pages} {134113} (\bibinfo {year} {2007})}\BibitemShut {NoStop}%
\bibitem [{\citenamefont {Jin}\ \emph {et~al.}(2008)\citenamefont {Jin},
  \citenamefont {Zheng},\ and\ \citenamefont {Yan}}]{jin2008-exact}%
  \BibitemOpen
  \bibfield  {author} {\bibinfo {author} {\bibfnamefont {J.}~\bibnamefont
  {Jin}}, \bibinfo {author} {\bibfnamefont {X.}~\bibnamefont {Zheng}},\ and\
  \bibinfo {author} {\bibfnamefont {Y.}~\bibnamefont {Yan}},\ }\bibfield
  {title} {\bibinfo {title} {Exact dynamics of dissipative electronic systems
  and quantum transport: Hierarchical equations of motion approach},\ }\href
  {https://doi.org/10.1063/1.2938087} {\bibfield  {journal} {\bibinfo
  {journal} {J. Chem. Phys.}\ }\textbf {\bibinfo {volume} {128}},\ \bibinfo
  {pages} {234703} (\bibinfo {year} {2008})}\BibitemShut {NoStop}%
\bibitem [{\citenamefont {Yan}\ \emph {et~al.}(2016)\citenamefont {Yan},
  \citenamefont {Jin}, \citenamefont {Xu},\ and\ \citenamefont
  {Zheng}}]{yan2016-dissipation}%
  \BibitemOpen
  \bibfield  {author} {\bibinfo {author} {\bibfnamefont {Y.}~\bibnamefont
  {Yan}}, \bibinfo {author} {\bibfnamefont {J.}~\bibnamefont {Jin}}, \bibinfo
  {author} {\bibfnamefont {R.-X.}\ \bibnamefont {Xu}},\ and\ \bibinfo {author}
  {\bibfnamefont {X.}~\bibnamefont {Zheng}},\ }\bibfield  {title} {\bibinfo
  {title} {Dissipation equation of motion approach to open quantum systems},\
  }\href {https://doi.org/10.1007/s11467-016-0513-5} {\bibfield  {journal}
  {\bibinfo  {journal} {Front. Phys.}\ }\textbf {\bibinfo {volume} {11}},\
  \bibinfo {pages} {110306} (\bibinfo {year} {2016})}\BibitemShut {NoStop}%
\bibitem [{\citenamefont {Cao}\ \emph {et~al.}(2023)\citenamefont {Cao},
  \citenamefont {Ye}, \citenamefont {Xu}, \citenamefont {Zheng},\ and\
  \citenamefont {Yan}}]{cao2023-recent}%
  \BibitemOpen
  \bibfield  {author} {\bibinfo {author} {\bibfnamefont {J.}~\bibnamefont
  {Cao}}, \bibinfo {author} {\bibfnamefont {L.}~\bibnamefont {Ye}}, \bibinfo
  {author} {\bibfnamefont {R.}~\bibnamefont {Xu}}, \bibinfo {author}
  {\bibfnamefont {X.}~\bibnamefont {Zheng}},\ and\ \bibinfo {author}
  {\bibfnamefont {Y.}~\bibnamefont {Yan}},\ }\bibfield  {title} {\bibinfo
  {title} {Recent advances in fermionic hierarchical equations of motion method
  for strongly correlated quantum impurity systems},\ }\href
  {https://doi.org/10.52396/justc-2022-0164} {\bibfield  {journal} {\bibinfo
  {journal} {JUSTC}\ }\textbf {\bibinfo {volume} {53}},\ \bibinfo {pages}
  {0302} (\bibinfo {year} {2023})}\BibitemShut {NoStop}%
\bibitem [{\citenamefont {Strand}\ \emph {et~al.}(2015)\citenamefont {Strand},
  \citenamefont {Eckstein},\ and\ \citenamefont {Werner}}]{strand2015-beyond}%
  \BibitemOpen
  \bibfield  {author} {\bibinfo {author} {\bibfnamefont {H.~U.~R.}\
  \bibnamefont {Strand}}, \bibinfo {author} {\bibfnamefont {M.}~\bibnamefont
  {Eckstein}},\ and\ \bibinfo {author} {\bibfnamefont {P.}~\bibnamefont
  {Werner}},\ }\bibfield  {title} {\bibinfo {title} {Beyond the hubbard bands
  in strongly correlated lattice bosons},\ }\href
  {https://doi.org/10.1103/physreva.92.063602} {\bibfield  {journal} {\bibinfo
  {journal} {Phys. Rev. A}\ }\textbf {\bibinfo {volume} {92}},\ \bibinfo
  {pages} {063602} (\bibinfo {year} {2015})}\BibitemShut {NoStop}%
\bibitem [{\citenamefont {Dong}\ \emph {et~al.}(2022)\citenamefont {Dong},
  \citenamefont {Gull},\ and\ \citenamefont {Strand}}]{dong2022-excitations}%
  \BibitemOpen
  \bibfield  {author} {\bibinfo {author} {\bibfnamefont {X.}~\bibnamefont
  {Dong}}, \bibinfo {author} {\bibfnamefont {E.}~\bibnamefont {Gull}},\ and\
  \bibinfo {author} {\bibfnamefont {H.~U.~R.}\ \bibnamefont {Strand}},\
  }\bibfield  {title} {\bibinfo {title} {Excitations and spectra from
  equilibrium real-time green’s functions},\ }\href
  {https://doi.org/10.1103/physrevb.106.125153} {\bibfield  {journal} {\bibinfo
   {journal} {Phys. Rev. B}\ }\textbf {\bibinfo {volume} {106}},\ \bibinfo
  {pages} {125153} (\bibinfo {year} {2022})}\BibitemShut {NoStop}%
\bibitem [{\citenamefont {Thoenniss}\ \emph
  {et~al.}(2023{\natexlab{a}})\citenamefont {Thoenniss}, \citenamefont
  {Lerose},\ and\ \citenamefont {Abanin}}]{ThoennissAbanin2023a}%
  \BibitemOpen
  \bibfield  {author} {\bibinfo {author} {\bibfnamefont {J.}~\bibnamefont
  {Thoenniss}}, \bibinfo {author} {\bibfnamefont {A.}~\bibnamefont {Lerose}},\
  and\ \bibinfo {author} {\bibfnamefont {D.~A.}\ \bibnamefont {Abanin}},\
  }\bibfield  {title} {\bibinfo {title} {Nonequilibrium quantum impurity
  problems via matrix-product states in the temporal domain},\ }\href
  {https://doi.org/10.1103/PhysRevB.107.195101} {\bibfield  {journal} {\bibinfo
   {journal} {Phys. Rev. B}\ }\textbf {\bibinfo {volume} {107}},\ \bibinfo
  {pages} {195101} (\bibinfo {year} {2023}{\natexlab{a}})}\BibitemShut
  {NoStop}%
\bibitem [{\citenamefont {Thoenniss}\ \emph
  {et~al.}(2023{\natexlab{b}})\citenamefont {Thoenniss}, \citenamefont
  {Sonner}, \citenamefont {Lerose},\ and\ \citenamefont
  {Abanin}}]{ThoennissAbanin2023b}%
  \BibitemOpen
  \bibfield  {author} {\bibinfo {author} {\bibfnamefont {J.}~\bibnamefont
  {Thoenniss}}, \bibinfo {author} {\bibfnamefont {M.}~\bibnamefont {Sonner}},
  \bibinfo {author} {\bibfnamefont {A.}~\bibnamefont {Lerose}},\ and\ \bibinfo
  {author} {\bibfnamefont {D.~A.}\ \bibnamefont {Abanin}},\ }\bibfield  {title}
  {\bibinfo {title} {Efficient method for quantum impurity problems out of
  equilibrium},\ }\href {https://doi.org/10.1103/PhysRevB.107.L201115}
  {\bibfield  {journal} {\bibinfo  {journal} {Phys. Rev. B}\ }\textbf {\bibinfo
  {volume} {107}},\ \bibinfo {pages} {L201115} (\bibinfo {year}
  {2023}{\natexlab{b}})}\BibitemShut {NoStop}%
\bibitem [{\citenamefont {Ng}\ \emph {et~al.}(2023)\citenamefont {Ng},
  \citenamefont {Park}, \citenamefont {Millis}, \citenamefont {Chan},\ and\
  \citenamefont {Reichman}}]{NgReichman2023}%
  \BibitemOpen
  \bibfield  {author} {\bibinfo {author} {\bibfnamefont {N.}~\bibnamefont
  {Ng}}, \bibinfo {author} {\bibfnamefont {G.}~\bibnamefont {Park}}, \bibinfo
  {author} {\bibfnamefont {A.~J.}\ \bibnamefont {Millis}}, \bibinfo {author}
  {\bibfnamefont {G.~K.-L.}\ \bibnamefont {Chan}},\ and\ \bibinfo {author}
  {\bibfnamefont {D.~R.}\ \bibnamefont {Reichman}},\ }\bibfield  {title}
  {\bibinfo {title} {Real-time evolution of anderson impurity models via tensor
  network influence functionals},\ }\href
  {https://doi.org/10.1103/PhysRevB.107.125103} {\bibfield  {journal} {\bibinfo
   {journal} {Phys. Rev. B}\ }\textbf {\bibinfo {volume} {107}},\ \bibinfo
  {pages} {125103} (\bibinfo {year} {2023})}\BibitemShut {NoStop}%
\bibitem [{\citenamefont {Chen}\ \emph
  {et~al.}(2024{\natexlab{a}})\citenamefont {Chen}, \citenamefont {Xu},\ and\
  \citenamefont {Guo}}]{ChenGuo2024a}%
  \BibitemOpen
  \bibfield  {author} {\bibinfo {author} {\bibfnamefont {R.}~\bibnamefont
  {Chen}}, \bibinfo {author} {\bibfnamefont {X.}~\bibnamefont {Xu}},\ and\
  \bibinfo {author} {\bibfnamefont {C.}~\bibnamefont {Guo}},\ }\bibfield
  {title} {\bibinfo {title} {Grassmann time-evolving matrix product operators
  for quantum impurity models},\ }\href
  {https://doi.org/10.1103/PhysRevB.109.045140} {\bibfield  {journal} {\bibinfo
   {journal} {Phys. Rev. B}\ }\textbf {\bibinfo {volume} {109}},\ \bibinfo
  {pages} {045140} (\bibinfo {year} {2024}{\natexlab{a}})}\BibitemShut
  {NoStop}%
\bibitem [{\citenamefont {Strathearn}\ \emph {et~al.}(2018)\citenamefont
  {Strathearn}, \citenamefont {Kirton}, \citenamefont {Kilda}, \citenamefont
  {Keeling},\ and\ \citenamefont {Lovett}}]{StrathearnLovett2018}%
  \BibitemOpen
  \bibfield  {author} {\bibinfo {author} {\bibfnamefont {A.}~\bibnamefont
  {Strathearn}}, \bibinfo {author} {\bibfnamefont {P.}~\bibnamefont {Kirton}},
  \bibinfo {author} {\bibfnamefont {D.}~\bibnamefont {Kilda}}, \bibinfo
  {author} {\bibfnamefont {J.}~\bibnamefont {Keeling}},\ and\ \bibinfo {author}
  {\bibfnamefont {B.~W.}\ \bibnamefont {Lovett}},\ }\bibfield  {title}
  {\bibinfo {title} {Efficient non-markovian quantum dynamics using
  time-evolving matrix product operators},\ }\href
  {https://doi.org/10.1038/s41467-018-05617-3} {\bibfield  {journal} {\bibinfo
  {journal} {Nat. Commun.}\ }\textbf {\bibinfo {volume} {9}},\ \bibinfo {pages}
  {3322} (\bibinfo {year} {2018})}\BibitemShut {NoStop}%
\bibitem [{\citenamefont {Chen}\ \emph
  {et~al.}(2024{\natexlab{b}})\citenamefont {Chen}, \citenamefont {Xu},\ and\
  \citenamefont {Guo}}]{ChenGuo2024c}%
  \BibitemOpen
  \bibfield  {author} {\bibinfo {author} {\bibfnamefont {R.}~\bibnamefont
  {Chen}}, \bibinfo {author} {\bibfnamefont {X.}~\bibnamefont {Xu}},\ and\
  \bibinfo {author} {\bibfnamefont {C.}~\bibnamefont {Guo}},\ }\bibfield
  {title} {\bibinfo {title} {Real-time impurity solver using grassmann
  time-evolving matrix product operators},\ }\href
  {https://doi.org/10.1103/PhysRevB.109.165113} {\bibfield  {journal} {\bibinfo
   {journal} {Phys. Rev. B}\ }\textbf {\bibinfo {volume} {109}},\ \bibinfo
  {pages} {165113} (\bibinfo {year} {2024}{\natexlab{b}})}\BibitemShut
  {NoStop}%
\bibitem [{\citenamefont {Guo}\ and\ \citenamefont
  {Chen}(2024{\natexlab{a}})}]{GuoChen2024d}%
  \BibitemOpen
  \bibfield  {author} {\bibinfo {author} {\bibfnamefont {C.}~\bibnamefont
  {Guo}}\ and\ \bibinfo {author} {\bibfnamefont {R.}~\bibnamefont {Chen}},\
  }\bibfield  {title} {\bibinfo {title} {{Efficient construction of the
  Feynman-Vernon influence functional as matrix product states}},\ }\href
  {https://doi.org/10.21468/SciPostPhysCore.7.3.063} {\bibfield  {journal}
  {\bibinfo  {journal} {SciPost Phys. Core}\ }\textbf {\bibinfo {volume} {7}},\
  \bibinfo {pages} {063} (\bibinfo {year} {2024}{\natexlab{a}})}\BibitemShut
  {NoStop}%
\bibitem [{\citenamefont {Kloss}\ \emph {et~al.}(2023)\citenamefont {Kloss},
  \citenamefont {Thoenniss}, \citenamefont {Sonner}, \citenamefont {Lerose},
  \citenamefont {Fishman}, \citenamefont {Stoudenmire}, \citenamefont
  {Parcollet}, \citenamefont {Georges},\ and\ \citenamefont
  {Abanin}}]{KlossAbanin2023}%
  \BibitemOpen
  \bibfield  {author} {\bibinfo {author} {\bibfnamefont {B.}~\bibnamefont
  {Kloss}}, \bibinfo {author} {\bibfnamefont {J.}~\bibnamefont {Thoenniss}},
  \bibinfo {author} {\bibfnamefont {M.}~\bibnamefont {Sonner}}, \bibinfo
  {author} {\bibfnamefont {A.}~\bibnamefont {Lerose}}, \bibinfo {author}
  {\bibfnamefont {M.~T.}\ \bibnamefont {Fishman}}, \bibinfo {author}
  {\bibfnamefont {E.~M.}\ \bibnamefont {Stoudenmire}}, \bibinfo {author}
  {\bibfnamefont {O.}~\bibnamefont {Parcollet}}, \bibinfo {author}
  {\bibfnamefont {A.}~\bibnamefont {Georges}},\ and\ \bibinfo {author}
  {\bibfnamefont {D.~A.}\ \bibnamefont {Abanin}},\ }\bibfield  {title}
  {\bibinfo {title} {Equilibrium quantum impurity problems via matrix product
  state encoding of the retarded action},\ }\href
  {https://doi.org/10.1103/physrevb.108.205110} {\bibfield  {journal} {\bibinfo
   {journal} {Phys. Rev. B}\ }\textbf {\bibinfo {volume} {108}},\ \bibinfo
  {pages} {205110} (\bibinfo {year} {2023})}\BibitemShut {NoStop}%
\bibitem [{\citenamefont {Chen}\ \emph
  {et~al.}(2024{\natexlab{c}})\citenamefont {Chen}, \citenamefont {Xu},\ and\
  \citenamefont {Guo}}]{ChenGuo2024b}%
  \BibitemOpen
  \bibfield  {author} {\bibinfo {author} {\bibfnamefont {R.}~\bibnamefont
  {Chen}}, \bibinfo {author} {\bibfnamefont {X.}~\bibnamefont {Xu}},\ and\
  \bibinfo {author} {\bibfnamefont {C.}~\bibnamefont {Guo}},\ }\bibfield
  {title} {\bibinfo {title} {Grassmann time-evolving matrix product operators
  for equilibrium quantum impurity problems},\ }\href
  {https://doi.org/10.1088/1367-2630/ad19fa} {\bibfield  {journal} {\bibinfo
  {journal} {New J. Phys.}\ }\textbf {\bibinfo {volume} {26}},\ \bibinfo
  {pages} {013019} (\bibinfo {year} {2024}{\natexlab{c}})}\BibitemShut
  {NoStop}%
\bibitem [{\citenamefont {Jarrell}\ and\ \citenamefont
  {Biham}(1989)}]{jarrell1989-dynamical}%
  \BibitemOpen
  \bibfield  {author} {\bibinfo {author} {\bibfnamefont {M.}~\bibnamefont
  {Jarrell}}\ and\ \bibinfo {author} {\bibfnamefont {O.}~\bibnamefont
  {Biham}},\ }\bibfield  {title} {\bibinfo {title} {Dynamical approach to
  analytic continuation of quantum monte carlo data},\ }\href
  {https://doi.org/10.1103/physrevlett.63.2504} {\bibfield  {journal} {\bibinfo
   {journal} {Phys. Rev. Lett.}\ }\textbf {\bibinfo {volume} {63}},\ \bibinfo
  {pages} {2504} (\bibinfo {year} {1989})}\BibitemShut {NoStop}%
\bibitem [{\citenamefont {Fei}\ \emph {et~al.}(2021)\citenamefont {Fei},
  \citenamefont {Yeh},\ and\ \citenamefont {Gull}}]{FeiGull2021}%
  \BibitemOpen
  \bibfield  {author} {\bibinfo {author} {\bibfnamefont {J.}~\bibnamefont
  {Fei}}, \bibinfo {author} {\bibfnamefont {C.-N.}\ \bibnamefont {Yeh}},\ and\
  \bibinfo {author} {\bibfnamefont {E.}~\bibnamefont {Gull}},\ }\bibfield
  {title} {\bibinfo {title} {Nevanlinna analytical continuation},\ }\href
  {https://doi.org/10.1103/PhysRevLett.126.056402} {\bibfield  {journal}
  {\bibinfo  {journal} {Phys. Rev. Lett.}\ }\textbf {\bibinfo {volume} {126}},\
  \bibinfo {pages} {056402} (\bibinfo {year} {2021})}\BibitemShut {NoStop}%
\bibitem [{\citenamefont {Guo}\ and\ \citenamefont
  {Chen}(2024{\natexlab{b}})}]{GuoChen2024e}%
  \BibitemOpen
  \bibfield  {author} {\bibinfo {author} {\bibfnamefont {C.}~\bibnamefont
  {Guo}}\ and\ \bibinfo {author} {\bibfnamefont {R.}~\bibnamefont {Chen}},\
  }\bibfield  {title} {\bibinfo {title} {Infinite grassmann time-evolving
  matrix product operator method in the steady state},\ }\href
  {https://doi.org/10.1103/physrevb.110.045106} {\bibfield  {journal} {\bibinfo
   {journal} {Phys. Rev. B}\ }\textbf {\bibinfo {volume} {110}},\ \bibinfo
  {pages} {045106} (\bibinfo {year} {2024}{\natexlab{b}})}\BibitemShut
  {NoStop}%
\bibitem [{\citenamefont {Kadanoff}\ and\ \citenamefont
  {Baym}(1962)}]{kadanoff1962-quantum}%
  \BibitemOpen
  \bibfield  {author} {\bibinfo {author} {\bibfnamefont {L.~P.}\ \bibnamefont
  {Kadanoff}}\ and\ \bibinfo {author} {\bibfnamefont {G.}~\bibnamefont
  {Baym}},\ }\href@noop {} {\emph {\bibinfo {title} {Quantum Statistical
  Mechnics}}}\ (\bibinfo  {publisher} {W. A. Benjamin},\ \bibinfo {address}
  {New York},\ \bibinfo {year} {1962})\BibitemShut {NoStop}%
\bibitem [{\citenamefont {Aoki}\ \emph {et~al.}(2014)\citenamefont {Aoki},
  \citenamefont {Tsuji}, \citenamefont {Eckstein}, \citenamefont {Kollar},
  \citenamefont {Oka},\ and\ \citenamefont {Werner}}]{AokiWerner2014}%
  \BibitemOpen
  \bibfield  {author} {\bibinfo {author} {\bibfnamefont {H.}~\bibnamefont
  {Aoki}}, \bibinfo {author} {\bibfnamefont {N.}~\bibnamefont {Tsuji}},
  \bibinfo {author} {\bibfnamefont {M.}~\bibnamefont {Eckstein}}, \bibinfo
  {author} {\bibfnamefont {M.}~\bibnamefont {Kollar}}, \bibinfo {author}
  {\bibfnamefont {T.}~\bibnamefont {Oka}},\ and\ \bibinfo {author}
  {\bibfnamefont {P.}~\bibnamefont {Werner}},\ }\bibfield  {title} {\bibinfo
  {title} {Nonequilibrium dynamical mean-field theory and its applications},\
  }\href {https://doi.org/10.1103/RevModPhys.86.779} {\bibfield  {journal}
  {\bibinfo  {journal} {Rev. Mod. Phys.}\ }\textbf {\bibinfo {volume} {86}},\
  \bibinfo {pages} {779} (\bibinfo {year} {2014})}\BibitemShut {NoStop}%
\bibitem [{\citenamefont {Makarov}\ and\ \citenamefont
  {Makri}(1994)}]{makarov1994-path}%
  \BibitemOpen
  \bibfield  {author} {\bibinfo {author} {\bibfnamefont {D.~E.}\ \bibnamefont
  {Makarov}}\ and\ \bibinfo {author} {\bibfnamefont {N.}~\bibnamefont
  {Makri}},\ }\bibfield  {title} {\bibinfo {title} {Path integrals for
  dissipative systems by tensor multiplication. condensed phase quantum
  dynamics for arbitrarily long time},\ }\href
  {https://doi.org/10.1016/0009-2614(94)00275-4} {\bibfield  {journal}
  {\bibinfo  {journal} {Chem. Phys. Lett.}\ }\textbf {\bibinfo {volume}
  {221}},\ \bibinfo {pages} {482} (\bibinfo {year} {1994})}\BibitemShut
  {NoStop}%
\bibitem [{\citenamefont {Makri}(1995)}]{makri1995-numerical}%
  \BibitemOpen
  \bibfield  {author} {\bibinfo {author} {\bibfnamefont {N.}~\bibnamefont
  {Makri}},\ }\bibfield  {title} {\bibinfo {title} {Numerical path integral
  techniques for long time dynamics of quantum dissipative systems},\ }\href
  {https://doi.org/10.1063/1.531046} {\bibfield  {journal} {\bibinfo  {journal}
  {J. Math. Phys.}\ }\textbf {\bibinfo {volume} {36}},\ \bibinfo {pages} {2430}
  (\bibinfo {year} {1995})}\BibitemShut {NoStop}%
\bibitem [{\citenamefont {Negele}\ and\ \citenamefont
  {Orland}(1998)}]{negele1998-quantum}%
  \BibitemOpen
  \bibfield  {author} {\bibinfo {author} {\bibfnamefont {J.~W.}\ \bibnamefont
  {Negele}}\ and\ \bibinfo {author} {\bibfnamefont {H.}~\bibnamefont
  {Orland}},\ }\href@noop {} {\emph {\bibinfo {title} {Quantum Many-Particle
  Systems}}}\ (\bibinfo  {publisher} {Westview Press},\ \bibinfo {year}
  {1998})\BibitemShut {NoStop}%
\bibitem [{\citenamefont {Leggett}\ \emph {et~al.}(1987)\citenamefont
  {Leggett}, \citenamefont {Chakravarty}, \citenamefont {Dorsey}, \citenamefont
  {Fisher}, \citenamefont {Garg},\ and\ \citenamefont
  {Zwerger}}]{LeggettZwerger1987}%
  \BibitemOpen
  \bibfield  {author} {\bibinfo {author} {\bibfnamefont {A.~J.}\ \bibnamefont
  {Leggett}}, \bibinfo {author} {\bibfnamefont {S.}~\bibnamefont
  {Chakravarty}}, \bibinfo {author} {\bibfnamefont {A.~T.}\ \bibnamefont
  {Dorsey}}, \bibinfo {author} {\bibfnamefont {M.~P.~A.}\ \bibnamefont
  {Fisher}}, \bibinfo {author} {\bibfnamefont {A.}~\bibnamefont {Garg}},\ and\
  \bibinfo {author} {\bibfnamefont {W.}~\bibnamefont {Zwerger}},\ }\bibfield
  {title} {\bibinfo {title} {Dynamics of the dissipative two-state system},\
  }\href {https://doi.org/10.1103/RevModPhys.59.1} {\bibfield  {journal}
  {\bibinfo  {journal} {Rev. Mod. Phys.}\ }\textbf {\bibinfo {volume} {59}},\
  \bibinfo {pages} {1} (\bibinfo {year} {1987})}\BibitemShut {NoStop}%
\bibitem [{\citenamefont {Lifshitz}\ and\ \citenamefont
  {Pitaevskii}(1980)}]{lifshitz1980-statistical}%
  \BibitemOpen
  \bibfield  {author} {\bibinfo {author} {\bibfnamefont {E.~M.}\ \bibnamefont
  {Lifshitz}}\ and\ \bibinfo {author} {\bibfnamefont {L.~P.}\ \bibnamefont
  {Pitaevskii}},\ }\href@noop {} {\emph {\bibinfo {title} {Course of
  Theoretical Physics Volume 9: Statistical Physics part 2}}}\ (\bibinfo
  {publisher} {Elsevier},\ \bibinfo {year} {1980})\BibitemShut {NoStop}%
\bibitem [{\citenamefont {Kaye}\ \emph {et~al.}(2022)\citenamefont {Kaye},
  \citenamefont {Chen},\ and\ \citenamefont {Strand}}]{kaye2022-libdlr}%
  \BibitemOpen
  \bibfield  {author} {\bibinfo {author} {\bibfnamefont {J.}~\bibnamefont
  {Kaye}}, \bibinfo {author} {\bibfnamefont {K.}~\bibnamefont {Chen}},\ and\
  \bibinfo {author} {\bibfnamefont {H.~U.}\ \bibnamefont {Strand}},\ }\bibfield
   {title} {\bibinfo {title} {libdlr: Efficient imaginary time calculations
  using the discrete lehmann representation},\ }\href
  {https://doi.org/10.1016/j.cpc.2022.108458} {\bibfield  {journal} {\bibinfo
  {journal} {Comput. Phys. Commun.}\ }\textbf {\bibinfo {volume} {280}},\
  \bibinfo {pages} {108458} (\bibinfo {year} {2022})}\BibitemShut {NoStop}%
\bibitem [{\citenamefont {Kaye}\ and\ \citenamefont
  {U.~R.~Strand}(2023)}]{kaye2023-fast}%
  \BibitemOpen
  \bibfield  {author} {\bibinfo {author} {\bibfnamefont {J.}~\bibnamefont
  {Kaye}}\ and\ \bibinfo {author} {\bibfnamefont {H.}~\bibnamefont
  {U.~R.~Strand}},\ }\bibfield  {title} {\bibinfo {title} {A fast time domain
  solver for the equilibrium dyson equation},\ }\href
  {https://doi.org/10.1007/s10444-023-10067-7} {\bibfield  {journal} {\bibinfo
  {journal} {Adv. Comput. Math.}\ }\textbf {\bibinfo {volume} {49}},\ \bibinfo
  {pages} {63} (\bibinfo {year} {2023})}\BibitemShut {NoStop}%
\end{thebibliography}%

\end{document}